\documentclass[11pt]{revtex4}

\usepackage{hyperref}
\usepackage{graphicx}% Include figure files
\usepackage{dcolumn}% Align table columns on decimal point
\usepackage{color}
\usepackage{bm}% bold math
\hypersetup{dvips,dvipdfm,colorlinks=true,urlcolor=magenta,filecolor=magenta,linktoc=page,citecolor=red,linkcolor=blue,bookmarks=true}
\usepackage{amsmath}
\usepackage{amsfonts}
\usepackage{amssymb}
\usepackage{epstopdf}
\usepackage{float}

\input epsf

\begin{document}

\title{Dynamical System Analysis of a Dirac-Born-Infeld model : A center Manifold perspective}

\author{Subhajyoti Pal\footnote {palsubhajyoti@gmail.com}}
\affiliation{Department of Mathematics, Sister Nibedita Govt
General Degree College For Girls, Kolkata-700027, West Bengal, India.}

\author{Subenoy Chakraborty\footnote {schakraborty.math@gmail.com}}
\affiliation{Department of Mathematics, Jadavpur University, Kolkata-700032, West Bengal, India.}

%%%%%%%%%%%%%%%%%%%%%%%%%%%%%%%%%%%%%%%%%%%%%%%%%%%%%%%%%%%%%%%%%%%%%%%%%%%%%%%%%%%%%%%%%%%%%%%%%%%%%%%%%%%%%%%%%%%%%%%%%%%%

\begin{abstract}

\noindent In this paper we present the cosmological dynamics of a perfect fluid and the Dark Energy (DE) component of the universe, where our model of the dark energy is the string-theoritic Dirac-Born-Infeld (DBI) model. We assume that the potential of the scalar field and the warp factor of the warped throat region of the compact space in the extra dimension for the DBI model are both exponential in nature. In the background of spatially flat Friedman-Robertson-Walker-Lema\^{\i}tre universe, the Einstein field equations for the DBI dark energy reduce to a system of autonomous dynamical system. We then perform a dynamical system analysis for this system. Our analysis is motivated by the invariant manifold approach of the mathematical dynamics. In this method, it is possible to reach a definite conclusion even when the critical points of a dynamical system are non-hyperbolic in nature. Since we find the complete set of critical points for this system, the center manifold analysis ensures that our investigation of this model leaves no stone unturned. We find some intersting results such as that for some critical points there are situations where scaling solutions exist. Finally we present various topologically different phase planes and stability diagrams and discuss the corresponding cosmological scenario.\\\\
{\bf Keywords\,:} Non-hyperbolic point, Center manifold, Field equations, DBI Field.\\\\
PACS Numbers\,: 98.80.-k, 05.45.-a, 02.40.Sf, 02.40.Tt\\\\

\end{abstract}

\maketitle

%\myclassification{04.70.Dy $-$ 04.60.Kz $-$ Black hole physics~~;\\$~~~~~~~~~~~~~$ 05.70.-a $-$ Thermodynamics~~;\\$~~~~~~~~~~~~~$  95.30.Tg $-$ Thermodynamic processes, equation of state~~;\\$~~~~~~~~~~~~~$  95.30.Sf $-$ Relativity and gravitation }

%\tableofcontents

%\newpage
%%%%%%%%%%%%%%%%%%%%%%%%%%%%%%%%%%%%%%%%%%%%%%%%%%%%%%%%%%%%%%%%%%%%%%%%%%%%%%%%%%%%%%%%%%%%%%%%%%%%%%%%%%%%%%%%%%%%%%%%%%%%
%%%%%%%%%%%%%%%%%%%%%%%%%%%%%%%%%%%%%%%%%%%%%%%%%%%%%%%%%%%%%%%%%%%%%%%%%%%%%%%%%%%%%%%%%%%%%%%%%%%%%%%%%%%%%%%%%%%%%%%%%%%%

%%%%%%%%%%%%%%%%%%%%%%%%%%%%%%%%%%%%%%%%%%%%%%%%%%%%%%%%%%%%%%%%%%%%%%%%%%
\section{Introduction} \label{intro}
%%%%%%%%%%%%%%%%%%%%%%%%%%%%%%%%%%%%%%%%%%%%%%%%%%%%%%%%%%%%%%%%%%%%%%%%%%

\noindent There are many evidences arised by analyzing the data from
various physical observations such as the Type Ia Supernova~\cite {Rei98,Perl99},
CMB anisotropies~\cite {Sper07,Komat09} and the Baryon Acoustic
Oscillations~\cite {Per07} to suggest that we are in a spatially
flat universe which is presently going through a phase of accelerated expansion.

\smallskip
\noindent This accelerated expansion has been theoretically
attributed to an unknown matter with huge negative
pressure called the dark energy. Although very little is known about the properties of the dark energy~\cite {AT10,ENO05,ST04,GSY12,PW09}, the natural candidate for it is the Cosmological Constant $\Lambda$~\cite {Wein89,Carrol01}. But the coincidence problem and the fine tuning problem~\cite {Wein89,AMS00,CKW03,S02,CLW98,SS00} are undesirable and unavoidable issues which arise with this choice.
To avoid these situations, lots of theories with dynamical dark energy models have been suggested. Among them, some scalar field
models like the Quintessence~\cite {RP88,CDS98,ZWS99,OAP05,OAP06}, \emph{K}-essence~\cite {AMS01,ADM99}, Phantom~\cite {C02,NOT05}, Dilatonic ghost condensate~\cite {PT04} and the Tachyonic models~\cite {P02} have drawn a lot of attention.

\smallskip
\noindent On the other hand, one very interesting model from the string theory which can describe the early accelerated expansion
is the Dirac-Born-Infeld (DBI) model~\cite {C05,CLR11,KSSY12,MC15,CSS10}. Here the motion of a D3-brane in the warped throat region of the compact space in the extra dimension causes the inflation. Due to the DBI action, the kinetic term is non-canonical, whereas its potential arises from the internal tensions between D-branes which encodes the geometry of the warped throat region of the compact space. Because of these facts, the DBI model is quite different from the usual slow-roll models of inflation~\cite {RP88} except for some of the simplest possible cases. We will see later that by constraining some parameters which arise in the evolution equations such as the "warp factor" we can get scaling solutions where the constraints determine some parameters related with the compactifation.

\smallskip
\noindent In this paper, we will try to find that whether the DBI model can be a successful candidate
for the DE such that it can explain the late time acceleration.
In this perspective, we derive the
Einstein's field equations for the
DBI field. The potential and the warp factor have been assumed to take an exponential form. 
Then by a certain change of variables, these
equations convert to an autonomous dynamical system. After that we find the complete set of critical points. Then some novel ideas from
the dynamical systems, namely the Invariant Manifold and the
Center Manifold theories ~\cite {Per91,AP90,Wig03,ASB,HS74,BCL12} are applied to compute
the center manifolds at all the non-hyperbolic critical points. With the help of this machinary we try to find the stability conditions for them. We consider all the theoretically valid values of the parameters of the autonomous system.
 We then present the stability diagrams for each family of the critical points and as many as possible topologically different phase plane diagrams.

\smallskip
\noindent We use this idea because,
to the best of our knowledge, all dynamical system analysis found in literature has one important shortcoming that
they fail to analyze the non-hyperbolic critical points. So a theory such as this would hopefully help us to find the answer for all possible cases and hopefully it will help us to find some new attractor or scaling solutions too.  

\smallskip
\noindent The organization of this article is as follows : The section
\ref{secI} describes the Einstein field equations and energy conservation relations for the DBI
model. Section \ref{secII} explains the change of variables which facilitate the formation of an
autonomous system. Section \ref{secIII} discusses the stability analysis of the critical points. This
section also describes the stability diagrams and the phase plane diagrams for different
topological scenario. A comparison between the DBI field with exponential potential and warp factor and athe quintessence model with exponential potential is presented at the end of this section. Lastly section \ref{secIV} presents the
cosmological interpretation of our results and concludes our
work.

%%%%%%%%%%%%%%%%%%%%%%%%%%%%%%%%%%%%%%%%%%%%%%%%%%%%%%%%%%%%%%%%%%%%%%%%%%
\section{Equations}\label{secI}
%%%%%%%%%%%%%%%%%%%%%%%%%%%%%%%%%%%%%%%%%%%%%%%%%%%%%%%%%%%%%%%%%%%%%%%%%%

\noindent The background of our model  is  the homogeneous
 and isotropic flat Friedman-Robertson-Walker-Lema\^{\i}tre spacetime with scale factor $a(t)$.
The universe is assumed to be filled up with non-interacting dark energy (DE) 
and dark matter (DM). Dark matter is idealized to be a perfect fluid with energy density $\rho_m$ and the
dark energy is assumed to be the DBI field $\phi$ with the
potential $V(\phi)$. 

\smallskip
\noindent The action of the DBI field $\phi$ has the following form :
\begin{equation}\label{dbi} S=-\int d^4x\sqrt{-g}(\frac{1}{f(\phi)}(\sqrt{1-2f(\phi)X}-1)-V(\phi)),\end{equation} where 
\begin{equation}\label{x}X=-\frac{1}{2}g^{\mu\nu}\partial_\mu\phi\partial_\nu\phi.\end{equation}

\noindent The potential $V(\phi)$ arises from the quantum interactions between the D3 brane specifying the DBI field and the other 
D-branes. $f(\phi)$ is the "warp factor" representing the inverse of the D3-brane tensions. In this paper, we assume that both
the potential and the warp factor are positive and have exponential forms. 

\smallskip
\noindent It can be shown that the energy density $\rho_\phi$ and the pressure
$p_\phi$ of the scalar field have the following expressions :

\begin{equation}\label{de.d} \rho_\phi=\frac{\nu^2}{\nu+1}\dot{\phi}^{2}+V(\phi)\end{equation}
and \begin{equation}\label{de.p}
p_\phi=\frac{\nu}{\nu+1}\dot{\phi}^{2}-V(\phi).\end{equation}

\noindent Where $\dot{}$ denotes differentiation with respect to cosmic time
$t$ and $\nu$ is similar to the Leorentz boost factor with the following form :

\begin{equation}\label{lorentzlike}\nu=\frac{1}{\sqrt{1-f(\phi)\dot{\phi}^2}}.\end{equation}
\noindent 
The dark matter is assumed to be a perfect fluid with the linear
equation of state
\begin{equation}\label{bm.pd}p_m= \omega_m\rho_m.\end{equation} where $p_m$ and $\rho_m$ are the pressure and the density of the fluid and $\omega_m$
is the adiabatic index of the DM.

\noindent The field equations for this model are
\begin{equation}\label{efe}3H^2=(\rho_m+\rho_\phi)\end{equation}
and
\begin{equation}\label{2ndefe} \dot{H}=-\frac{1}{2}[\nu\dot{\phi}+\rho_m(1+\omega_m)].\end{equation} where $H=\frac{\dot{a}}{a}$ is the Hubble parameter.

\noindent We define the density parameters for the DE and the DM as $\Omega_\phi=\frac{\rho_\phi}{3H^2}$ and $\Omega_m=\frac{\rho_m}{3H^2}.$ Then it is clear that $\Omega_\phi+\Omega_m=1.$

\noindent The energy conservation relations are the following :
\begin{equation}\label{ecedm}\dot{\rho_m}+3H(1+\omega_m)\rho_m=0\end{equation} and 
\begin{equation}\label{ecede}\dot{\rho_\phi}+3H(1+\omega_\phi)\rho_\phi=0.\end{equation}

\noindent Then from (\ref{de.d}),(\ref{de.p}),(\ref{efe}),(\ref{2ndefe}) and
(\ref{ecede}) we see that the DBI field statisfies the following equation :
\begin{equation}\label{efed}\frac{2\nu^2}{\nu+1}\dot{\phi}\ddot{\phi}+(\frac{2\nu}{\nu+1}-\frac{\nu^2}{(\nu+1)^2})\dot{\nu}\dot{\phi}^2+\frac{dV(\phi)}{d\phi}\dot{\phi}+3H\nu\dot{\phi}^2=0.\end{equation}

\noindent Where $\ddot{}$ is the double differentiation with respect to the cosmic time $t.$

\smallskip

\noindent Finally the dynamics of our model is governed by the equations (\ref{2ndefe}), (\ref{ecedm}) and (\ref{efed}).
Our next job is to find proper
coordinate changes such that these evolution equations form an
autonomous dynamical system. This is done in
section \ref{secII}.

%%%%%%%%%%%%%%%%%%%%%%%%%%%%%%%%%%%%%%%%%%%%%%%%%%%%%%%%%%%%%%%%%%%%%%%%%%
\section{The autonomous system}\label{secII}
%%%%%%%%%%%%%%%%%%%%%%%%%%%%%%%%%%%%%%%%%%%%%%%%%%%%%%%%%%%%%%%%%%%%%%%%%%

\noindent We introduce the following coordinate changes :

\begin{equation}\label{xchange} x= \frac{\nu\dot{\phi}}{\sqrt{3(1+\nu)}H},\end{equation} and
\begin{equation}\label{ychange} y= \frac{\sqrt{V(\phi)}}{\sqrt{3}H}.\end{equation}

\noindent These changes transform the evolution equations as the
following autonomous system :

\begin{eqnarray}\label{ads}\frac{dx}{dN}= -\lambda \frac{\sqrt{3(1+\nu)}}{2\nu}y^2-\frac{3x}{2}[\frac{(1-x^2)}{\nu}+y^2-\omega_m(1-x^2-y^2)]\\
\label{ads1}\frac{dy}{dN}= \lambda\frac{\sqrt{3(1+\nu)}}{2\nu} xy+\frac{3y}{2}[1+\frac{x^2}{\nu}-y^2+\omega_m(1-x^2-y^2)]\\
\label{ads2}\frac{d\nu}{dN}= 2(\nu-1)[\frac{\sqrt{3(1+\nu)}}{2\nu}(\mu x-\lambda\frac{y^2}{x})-\frac{3(1+\nu)}{2\nu}]\end{eqnarray}

\noindent where $\lambda=\frac{1}{V(\phi)}\frac{dV(\phi)}{d\phi},$ $\mu=\frac{1}{f(\phi)}\frac{df(\phi)}{d\phi}$ and $N$= ln $a(t).$

\noindent Since the potential and the warp factor has exponential forms, $\lambda$ and $\mu$ are parameters with real values.

\smallskip

\noindent  Next we find the relevant
cosmological parameters in terms of the above transformed
variables :

\begin{equation}\omega_\phi=\frac{p_\phi}{\rho_\phi}=\frac{\frac{x^2}{\nu}-y^2}{x^2+y^2},\end{equation}
\begin{equation}\omega_{eff}=\frac{p_\phi+p_m}{\rho_\phi+\rho_m}=\frac{x^2}{\nu}-y^2+\omega_m(1-x^2-y^2)\end{equation}
and
\begin{equation}\label{dece}q=-(1+\frac{\dot{H}}{H^2})= \frac{1}{2}+\frac{3x^2}{2\nu}-\frac{3y^2}{2}+\frac{3\omega_m}{2}-\frac{3\omega_mx^2}{2}-\frac{3\omega_my^2}{2}.\end{equation}

\noindent $\omega_\phi$ and $\omega_{eff}$ are the equation of state and the effective equation of state parameters of the DBI field respectively. $q$ is the deceleration parameter. For the accelerated expansion, it is necessary that $\omega_{eff}
< -\frac{1}{3}$ and $q< 0.$

\noindent Now we procced to the next section to find the critical points of the dynamical system (\ref{ads})-(\ref{ads2}) and do the stability analysis.

%%%%%%%%%%%%%%%%%%%%%%%%%%%%%%%%%%%%%%%%%%%%%%%%%%%%%%%%%%%%%%%%%%%%%%%%%%
\section{Stability Analysis}\label{secIII}
%%%%%%%%%%%%%%%%%%%%%%%%%%%%%%%%%%%%%%%%%%%%%%%%%%%%%%%%%%%%%%%%%%%%%%%%%%

\noindent  There are ten critical points of this autonomous system. It will be shown in the following sections that the critical points in pair namely $(C_1,C_2), (C_3,C_4), (C_5,C_6), (C_7,C_8)$ and $(C_9,C_{10})$ are identical from the stability analysis and the cosmological view point (see TABLE \ref{tablast}). Further, the critical points $C_1, C_3, C_5, C_7$ and $C_{10}$ are identical to the critical points $(b1), (c1), (c2), (b2)$ and $(b3)$ in the TABLE I in ~\cite{CSS10}. Note that the other critical points in ~\cite{CSS10}, namely $(a1)-(a5)$ are ultrarelativistic in nature and the physical system will no longer have DBI type field and and it is also true for the critical point $(c3).$ Hence we have not considered these critical points. Further, in ~\cite{CSS10} the critical points are chosen to be hyperbolic in nature by adjusting the parameters involved, so that the eigen values of the Jacobian matrix of the linearized system are all non-zero. But in the present work, we consider the critical points to be non-hyperbolic in nature by considering one or two eigen values to be zero. The critical points are listed in the TABLE \ref{tab1} with the corresponding restriction on the parameters. 
\noindent
\begin{table}[H]
\centering%
\caption{Critical Points}\label{tab1}
\bigskip 
\begin{tabular}{|c|c|c|c|c|}
  \hline
  Critical Point Name & Critical Point & $\omega_m$ & $\lambda$ & $\mu$ \\
  \hline
  $C_1$ & $(1,0,1)$ & NA & NA & NA \\
  \hline
  $C_2$ & $(-1,0,1)$ & NA & NA & NA \\
  \hline
  $C_3$ & $(1,0,\frac{\mu^2}{3}-1)$ & NA & NA & $\sqrt{6}< \mu$ \\
  \hline
  $C_4$ & $(-1,0,\frac{\mu^2}{3}-1)$ & NA & NA & $\mu< -\sqrt{6}$ \\
  \hline
  $C_5$ & $(\frac{\sqrt{3(1+\frac{1}{\omega_m})}}{\mu},0,\frac{1}{\omega_m})$ & $0 < \omega_m\leq 1$ & NA & $\mu\neq 0$
\\
  \hline
  $C_6$ & $(-\frac{\sqrt{3(1+\frac{1}{\omega_m})}}{\mu},0,\frac{1}{\omega_m})$ & $0 < \omega_m\leq 1$ & NA & $\mu\neq 0$ \\
  \hline
  $C_7$ & $(-\frac{\lambda}{\sqrt{6}},\frac{\sqrt{6-\lambda^2}}{\sqrt{6}},1)$ & NA & $\lambda\neq 0,|\lambda|<\sqrt{6}$ & NA  \\
  \hline
  $C_8$ & $(-\frac{\lambda}{\sqrt{6}},-\frac{\sqrt{6-\lambda^2}}{\sqrt{6}},1)$ & NA & $\lambda\neq 0,|\lambda|<\sqrt{6}$ & NA \\
  \hline
  $C_9$ & $(-\frac{\sqrt{6}(\omega_m+1)}{2\lambda}),\frac{\sqrt{6(1-\omega_m^2)}}{2\lambda},1)$ & $-1< \omega_m\leq 1$ &  $\lambda\neq 0$ & NA\\
  \hline
  $C_{10}$ & $(-\frac{\sqrt{6}(\omega_m+1)}{2\lambda}),-\frac{\sqrt{6(1-\omega_m^2)}}{2\lambda},1)$  & $-1< \omega_m\leq  1$ & $\lambda\neq 0$ & NA\\
  \hline
\end{tabular}
\end{table}

\bigskip

\noindent Throughout this article "NA" means Not Applicable. The value of the relevant cosmological parameters for
each of the critical points are given in the TABLE \ref{tablast}.

\begin{table}[H]
\centering%
\caption{Values of the Different Cosmological Parameters at the Critical
Points}\label{tablast}
\bigskip
\begin{tabular}{|c|c|c|c|c|c|}
  \hline
  Critical Points & $\omega_\phi$ & $\omega_{eff}$ & $\Omega_m$ & $\Omega_\phi$ & $q$ \\
  \hline
  $C_1$ & $1$ & $1$ & $0$ & $1$ & $2$ \\
  \hline
  $C_2$ & $1$ & $1$ & $0$ & $1$ & $2$ \\
  \hline
  $C_3$ & $\frac{3}{\mu^2-3}$ & $\frac{3}{\mu^2-3}$ & $0$ & $1$ & $\frac{1}{2}+\frac{9}{2(\mu^2-3)}$ \\
  \hline
  $C_4$ & $\frac{3}{\mu^2-3}$ & $\frac{3}{\mu^2-3}$ & $0$ & $1$ & $\frac{1}{2}+\frac{9}{2(\mu^2-3)}$ \\
  \hline
  $C_5$ & $\omega_m$ & $\omega_m$ & $1-\frac{3(1+\omega_m)}{\omega_m\mu^2}$ & $\frac{3(1+\omega_m)}{\omega_m\mu^2}$ & $\frac{1}{2}+\frac{3\omega_m}{2}$ \\
  \hline
  $C_6$ & $\omega_m$ & $\omega_m$ & $1-\frac{3(1+\omega_m)}{\omega_m\mu^2}$ & $\frac{3(1+\omega_m)}{\omega_m\mu^2}$ & $\frac{1}{2}+\frac{3\omega_m}{2}$ \\
  \hline
  $C_7$ & $\frac{\lambda^2}{3}-1$ & $\frac{\lambda^2}{3}-1$ & $0$ & $1$ & $\frac{\lambda^2}{2}-1$ \\
  \hline
  $C_8$ & $\frac{\lambda^2}{3}-1$ & $\frac{\lambda^2}{3}-1$ & $0$ & $1$ & $\frac{\lambda^2}{2}-1$ \\
  \hline
  $C_9$ & $\omega_m$ & $\omega_m$ & $1-\frac{3(1+\omega_m)}{\lambda^2}$ & $\frac{3(1+\omega_m)}{\lambda^2}$ & $\frac{1}{2}+\frac{3\omega_m}{2}$ \\
  \hline
  $C_{10}$ & $\omega_m$ & $\omega_m$ & $1-\frac{3(1+\omega_m)}{\lambda^2}$ & $\frac{3(1+\omega_m)}{\lambda^2}$ & $\frac{1}{2}+\frac{3\omega_m}{2}$ \\
  \hline
\end{tabular}
 \end{table}

\bigskip

\noindent To compute the center manifold and the reduced system for each subcase of every critical point, we will do two successive transformations in the following manner.
\noindent If

$X=\left(%
\begin{array}{c}
  x \\
  y \\
\nu \\
\end{array}%
\right),$  $\bar{X}=\left(%
\begin{array}{c}
  \bar{x} \\
  \bar{y} \\
\bar{\nu} \\
\end{array}%
\right)$ and $\bar{\bar{X}}=\left(%
\begin{array}{c}
  \bar{\bar{x}} \\
  \bar{\bar{y}}\\
\bar{\bar{\nu}}\\
\end{array}%
\right),$

\noindent

\noindent then $\bar{X}=X-A$ and $\bar{\bar{X}}=P^{-1}\bar{X}$
for some $3\times 1$ and $3\times 3$ matrices $A$ and $P$ respectively, where $P$ is non-singular. The exact expression of $A$
and $P$ will be different for each subcase. We will write them explicitly as they appear.

\smallskip

\noindent We will see that although in the beginning, the coefficients appearing in various equations in the subsequent subsections are relatively smaller in size to write down, as this article progresses further, they start to appear in gigantic sizes. Hence it is unavoidable that we have to introduce some notations to express them. If the expressions are short to write, we write them explicitly, we don't use any notation. Otherwise, we use the following notations.
\bigskip

\noindent $A^n$ is the $A$ for the $n$-th critical point. $a^n_i$  denotes the coefficient of $s^i$
of the characteristic polynomial associated with the jacobian matrix of the system at the $n$-th critical point, where $s$ is the indeterminate of the polynomial. $e^n_i$  denotes the $i$-th eigen value of the system for the $n$-th critical point. $P^n$ denotes the $P$ for the $n$-th critical point whereas by $P^n_i$ we will mean the explicit form of $P$ for the $i$-th subcase of the $n$-th critical point. 

\noindent By $C^{n,i}_{j,(k,l,m)}$ we will denote the coefficient of $x^ky^l\nu^m$ of the $j$-th equation of the system of equations representing the center manifold of the $i$-th subcase of the $n$-th critical point. Similarly, $R^{n,i}_{j,(k,l,m)}$  will denote the coefficient of $x^ky^l\nu^m$ of the $j$-th equation of the system of equations representing the reduced system of the $i$-th subcase of the $n$-th critical point. $'$ will represent derivative with respect to $N.$

\smallskip

\noindent As the results of the stability analysis for the hyperbolic cases are easy application of the analysis of the corresponding linear system by the celebrated "Hartman-Gr\"{o}bman Theorem"  and are found abundantly in literature~\cite {KSSY12,MC15,CSS10}, we will not repeat these in our paper. Instead, we will list all the results corresponding to the stability analysis of the non-hyperbolic points in tables. Although for the sake of completeness,  at the end of each subsection we will provide a color graph representing the stability of the system for different values of the cosmological parameters for both the hyperbolic and non-hyperbolic cases which arise with each of the ten families of critical points. It will be seen in the coming subsections that in the color graphs, the fourth mini-figure in a figure is obtained by superimposing the first, second and the third mini-figure to highlight the stable, unstable and saddle zones for the respective critical points.

\smallskip

\noindent In the following subsections, "RS" stands for the reduced system, "ND" stands for the word "Not Determined" and "ODE" represent the word Ordinary Differential Equation.

%%% ----------------------------------------------------------------------

\subsection{Critical Point $C_1$}

\noindent For the critical point $C_1$, $A^1=\left(%
\begin{array}{c}
  1 \\
  0 \\
  1 \\
\end{array}%
\right)$ and the Jacobian matrix of the system (\ref{ads})-(\ref{ads2}) at
this critical point has the characteristic polynomial
\begin{equation}a^1_3s^3+a^1_2s^2+a^1_1s+a^1_0=0\end{equation}
where \begin{equation}a^1_3=1, a^1_2= (3\omega_m-\frac{\sqrt{6}\lambda}{2}-\sqrt{6}\mu)\end{equation} 
and
\begin{equation}a^1_1= [\frac{(\sqrt{6}\mu-6)(\sqrt{6}\lambda-6\omega_m+12)}{2}-\frac{(3\omega_m-3)(\sqrt{6}\lambda+6)}{2}] , a^1_0=\frac{(3\omega_m-3)(\sqrt{6}\lambda+6)(\sqrt{6}\mu-6)}{2}.\end{equation}

\noindent This polynomial has eigen values as \begin{equation}e^1_1=3-3\omega_m, e^1_2=\frac{\sqrt{6}\lambda}{2}+3, e^1_3=\sqrt{6}\mu-6.\end{equation}

\noindent We note that for a valid situation where $\omega_m=1, \lambda= -\sqrt{6}$ and $\mu= \sqrt{6},$ the center manifold reduction fails as all the three eigen values are zero in this case.
\smallskip

\noindent We present the TABLE \ref{tab2} containing the
non-hyperbolic subcases and the result of their stability
analysis.

\bigskip
\begin{table}[H]
\centering%
\caption{$C_1$(Center Manifolds and Reduced System)}\label{tab2}
\bigskip
\begin{tabular}{|c|c|c|c|c|c|}
  \hline
 Case & $\omega_m$ & $\lambda$ & $\mu$ & Center Manifold & Reduced System \\
  \hline
  a & $1$ & NA & NA & {$\bar{\bar{y}}=O(\bar{\bar{x}}^3),\bar{\bar{\nu}}=O(\bar{\bar{x}}^3)$} & $\bar{\bar{x}}'= 0$\\
  \hline
   b & NA & $-\sqrt{6}$ & NA & $\bar{\bar{y}}=-\frac{1}{2}\bar{\bar{x}}^2+O(\bar{\bar{x}}^3),\bar{\bar{\nu}}=O(\bar{\bar{x}}^3)$ & $\bar{\bar{x}}'=-\frac{3}{2}\bar{\bar{x}}^3$ \\
  \hline
   c & NA & NA & $\sqrt{6}$ & $\bar{\bar{y}}=O(\bar{\bar{x}}^3),\bar{\bar{\nu}}=O(\bar{\bar{x}}^3)$ & $\bar{\bar{x}}'= -\frac{3}{2}\bar{\bar{x}}^2$  \\
  \hline
  d & $1$ & $-\sqrt{6}$ & NA & $\bar{\bar{\nu}}= O(||(\bar{\bar{x}},\bar{\bar{y}})||^3)$ & $\bar{\bar{x}}'=-3\bar{\bar{x}}\bar{\bar{y}}^2, \bar{\bar{y}}'=-3\bar{\bar{x}}\bar{\bar{y}}$\\
  \hline
  e & $1$ & NA & $\sqrt{6}$ & $\bar{\bar{\nu}}= O(||(\bar{\bar{x}},\bar{\bar{y}})||^3)$ & $\bar{\bar{x}}'=-3\bar{\bar{x}}\bar{\bar{y}}, \bar{\bar{y}}'= 6\bar{\bar{x}}\bar{\bar{y}}-\frac{3}{2}\bar{\bar{y}}^2$\\
  \hline
 f & NA & $-\sqrt{6}$ & $\sqrt{6}$ & $\bar{\bar{\nu}}= -\frac{1}{2}\bar{\bar{x}}^2+ O(||(\bar{\bar{x}},\bar{\bar{y}})||^3)$ & $\bar{\bar{x}}'=\frac{3}{4}\bar{\bar{x}}\bar{\bar{y}}, \bar{\bar{y}}'= -\frac{3}{2}\bar{\bar{y}}^2$\\
  \hline
\end{tabular}
\end{table}

\noindent Where $O$ represents higher degree terms. For example, $O(x^m)$ represents sum of the terms having degree more than or equal to $m.$ Since the center manifold and reduced system primarily depend upon the first nonzero nonlinear term, sometimes we will even omit the terms with higher degree than that. In particular, if the expression for the center manifold or reduced system is large, we will omit the $O$ terms.

\noindent For all these subcases,
$P^1_i=\left(%
\begin{array}{ccc}
  1 & 0 & 0 \\
  0 & 1 & 0 \\
  0 & 0 & 1 \\
\end{array}%
\right),$

\noindent where $i=a, b, c, d, e, f.$

\smallskip

\noindent We summarize our results for the stability of the reduced system of $C_1$ in the TABLE \ref{tab3}.
\bigskip
\begin{table}[H]
\centering%
\caption{Summary for the critical point $C_1$(non-hyperbolic cases)}\label{tab3}
\bigskip
\begin{tabular}{|c|c|c|c|c|}
  \hline
  Case & $\omega_m$ & $\lambda$ & $\mu$ & Stability(RS)\\
  \hline
  a & $1$ & NA & NA & ND \\
  \hline
  b & NA & $-\sqrt{6}$ & NA & Stable \\
  \hline
  c & NA & NA & $\sqrt{6}$ & Stable \\
  \hline
  d & $1$ & $-\sqrt{6}$ & NA & Stable\\
  \hline
  e & $1$ & NA & $\sqrt{6}$ &  Stable \\
  \hline
  f & NA & $-\sqrt{6}$ & $\sqrt{6}$ & Unstable \\
  \hline
\end{tabular}
\end{table}

\noindent In the subcase (a), the center manifold analysis fails to determine stability as the reduced system represents a center. Hence some higher dimensional analysis may be needed to determine the stability for this subcase.

\begin{figure}[H]
\begin{center}
\includegraphics
[scale=0.45]{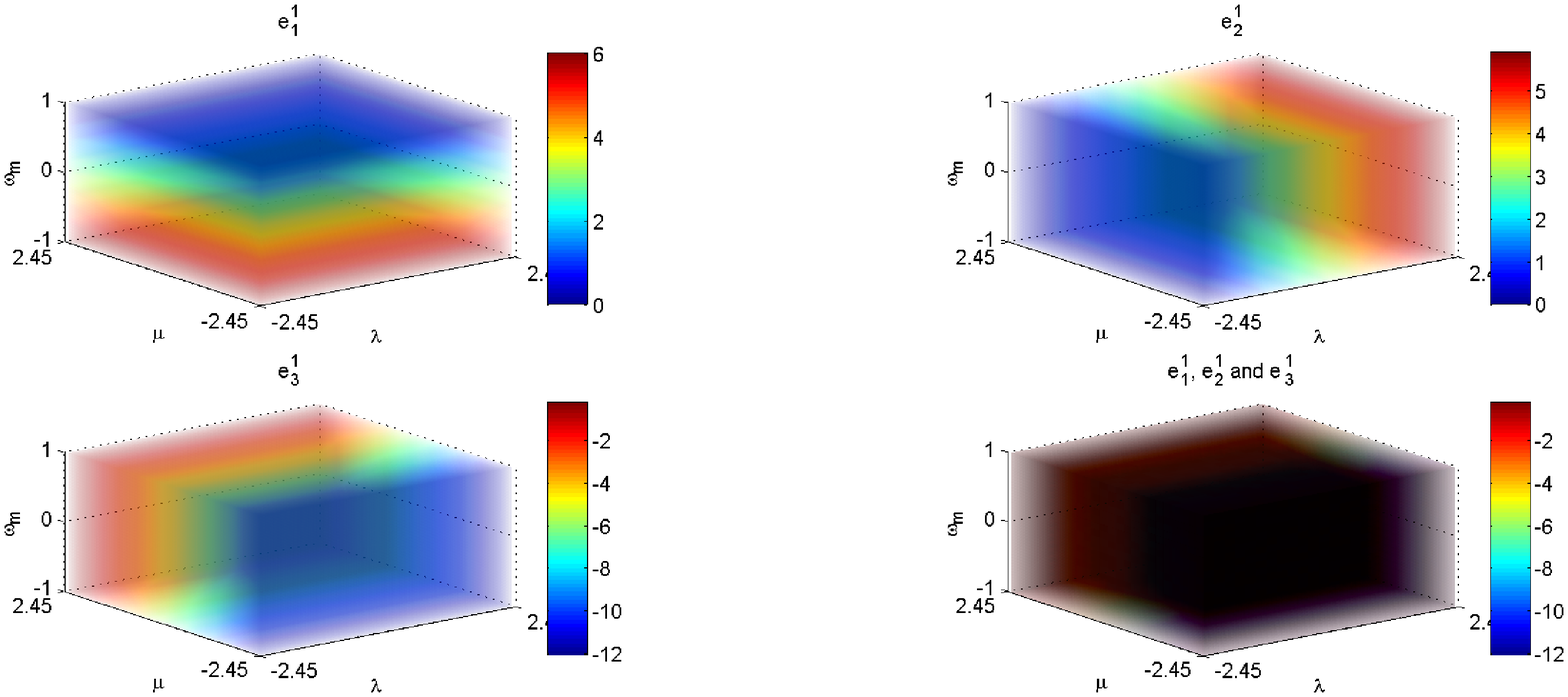}
\caption{$C_1$ }\label{fig101}
\end{center}
\end{figure}

\noindent The stability analysis for $C_1$ for different values of parameters in all the possible cases (both the hyperbolic and the non-hyperbolic) can be represented in the FIG.\ref{fig101}. In FIG.\ref{fig101}, the stability results of $C_1$ in the parameter region $\{-1\le\omega_m\le 1,-\sqrt{6}\le \lambda\le \sqrt{6},-\sqrt{6}\le \mu\le \sqrt{6}\}$ is presented.

\subsection{Critical Point $C_2$}

\noindent For critical point $C_2$, $A^2=\left(%
\begin{array}{c}
  -1 \\
  0 \\
  1 \\
\end{array}%
\right)$ and the Jacobian matrix of the system (\ref{ads})-(\ref{ads2}) at
this critical point has the characteristic polynomial
\begin{equation}a^2_3s^3+a^2_2s^2+a^2_1s+a^2_0=0\end{equation}
where \begin{equation}a^2_3=1, a^2_2= (-3\omega_m+\frac{\sqrt{6}\lambda}{2}+\sqrt{6}\mu+6)\end{equation} 
and
\begin{equation}a^2_1= [-\frac{(3\omega_m-3)(\sqrt{6}\lambda-6)}{2}+\frac{(6\omega_m-\sqrt{6}\lambda)(\sqrt{6}\mu+6)}{2}] , a^2_0=-\frac{(3\omega_m-3)(\sqrt{6}\lambda-6)(\sqrt{6}\mu+6)}{2}.\end{equation}
This has eigen values as \begin{equation}e^2_1= 3\omega_m-3, e^2_2= 3-\frac{\sqrt{6}\lambda}{2},
e^2_3= -\sqrt{6}\mu-6.\end{equation}  

\smallskip

\noindent We again note that for the situation where $\omega_m=1, \lambda= \sqrt{6}$ and $\mu=-\sqrt{6},$ the center manifold reduction fails as all the three eigenvalues are zero in this case too.
\smallskip

\noindent We present the TABLE \ref{tab4} containing the
non-hyperbolic subcases and the result of their stability
analysis :

\bigskip
\begin{table}[H]
\centering%
\caption{$C_2$(Center Manifolds and Reduced System)}\label{tab4}
\bigskip
\begin{tabular}{|c|c|c|c|c|c|}
  \hline
 Case & $\omega_m$ & $\lambda$ & $\mu$ & Center Manifold & Reduced System \\
  \hline
 a &  $1$ & NA & NA & {$\bar{\bar{y}}=O(\bar{\bar{x}}^3),\bar{\bar{\nu}}=O(\bar{\bar{x}}^3)$} & $\bar{\bar{x}}'= 0$\\
  \hline
 b &  NA & $\sqrt{6}$ & NA & $\bar{\bar{y}}=\frac{1}{2}\bar{\bar{x}}^2+O(\bar{\bar{x}}^3),\bar{\bar{\nu}}=O(\bar{\bar{x}}^3)$ & $\bar{\bar{x}}'=-\frac{3}{2}\bar{\bar{x}}^3$ \\
  \hline
 c & NA & NA & $-\sqrt{6}$ & $\bar{\bar{y}}=O(\bar{\bar{x}}^3),\bar{\bar{\nu}}=O(\bar{\bar{x}}^3)$ & $\bar{\bar{x}}'= -\frac{3}{2}\bar{\bar{x}}^2$  \\
  \hline
 d & $1$ & $\sqrt{6}$ & NA & $\bar{\bar{\nu}}= O(||(\bar{\bar{x}},\bar{\bar{y}})||^3)$ & $\bar{\bar{x}}'=-3\bar{\bar{x}}\bar{\bar{y}}^2, \bar{\bar{y}}'=3\bar{\bar{x}}\bar{\bar{y}}$\\
  \hline
 e & $1$ & NA & $-\sqrt{6}$ & $\bar{\bar{\nu}}= O(||(\bar{\bar{x}},\bar{\bar{y}})||^3)$ & $\bar{\bar{x}}'=3\bar{\bar{x}}\bar{\bar{y}}, \bar{\bar{y}}'=-6\bar{\bar{x}}\bar{\bar{y}}-\frac{3}{2}\bar{\bar{y}}^2$\\
  \hline
 f &  NA & $\sqrt{6}$ & $-\sqrt{6}$ & $\bar{\bar{\nu}}= \frac{1}{2}\bar{\bar{x}}^2+ O(||(\bar{\bar{x}},\bar{\bar{y}})||^3)$ & $\bar{\bar{x}}'=\frac{3}{4}\bar{\bar{x}}\bar{\bar{y}}, \bar{\bar{y}}'= -\frac{3}{2}\bar{\bar{y}}^2$\\
  \hline
\end{tabular}
\end{table}

For subcase (a), the $P^2_a$ is $\left(%
\begin{array}{ccc}
  1 & 0 & 0 \\
  0 & 1 & 0 \\
  0 & 0 & 1 \\
\end{array}%
\right).$

For subcase (b), $P^2_b= \left(%
\begin{array}{ccc}
  0 & 1 & 0 \\
  1 & 0 & 0 \\
  0 & 0 & 1 \\
\end{array}%
\right).$

For subcase (c), $P^2_c=\left(%
\begin{array}{ccc}
  0 & 1 & 0 \\
  0 & 0 & 1 \\
  1 & 0 & 0 \\
\end{array}%
\right).$

For subcase (d), $P^2_d= \left(%
\begin{array}{ccc}
  1 & 0 & 0 \\
  0 & 1 & 0 \\
  0 & 0 & 1 \\
\end{array}%
\right).$

For subcase (e), $P^2_e= \left(%
\begin{array}{ccc}
  1 & 0 & 0 \\
  0 & 0 & 1 \\
  0 & 1 & 0 \\
\end{array}%
\right).$

For subcase (f), $P^2_f= \left(%
\begin{array}{ccc}
  0 & 0 & 1 \\
  1 & 0 & 0 \\
  0 & 1 & 0 \\
\end{array}%
\right).$

\smallskip

We summarize our results for the stability of the reduced system of $C_2$ in the TABLE \ref{tab5}.

\bigskip

\begin{table}[H]
\centering%
\caption{Summary for the critical point $C_2$(non-hyperbolic cases)}\label{tab5}
\bigskip
\begin{tabular}{|c|c|c|c|c|}
  \hline
  Case & $\omega_m$ & $\lambda$ & $\mu$ & Stability(RS)\\
  \hline
  a & $1$ & NA & NA & ND \\
  \hline
  b & NA & $\sqrt{6}$ & NA & Stable \\
  \hline
  c & NA & NA & $-\sqrt{6}$ & Stable \\
  \hline
  d & $1$ & $\sqrt{6}$ & NA & Stable\\
  \hline
  e & $1$ & NA & $-\sqrt{6}$ & Unstable \\
  \hline
  f & NA & $\sqrt{6}$ & $-\sqrt{6}$ & Unstable \\
  \hline
\end{tabular}
\end{table}

\noindent Where we come to the conclusion as above by solving the ODE/pair of ODEs of the simplified reduced system. In the subcase (a), the center manifold analysis fails again to determine stability as the reduced system represents a center. Hence some higher dimensional analysis may be needed to determine the stability for this subcase. In the subcase (e), we found that the solution of the system of ODEs that govern the reduced dynamics for $C_2$ is sensitive to initial conditions, it is chaotic in nature indeed.

\begin{figure}[H]
\begin{center}
\includegraphics
[scale=0.45]{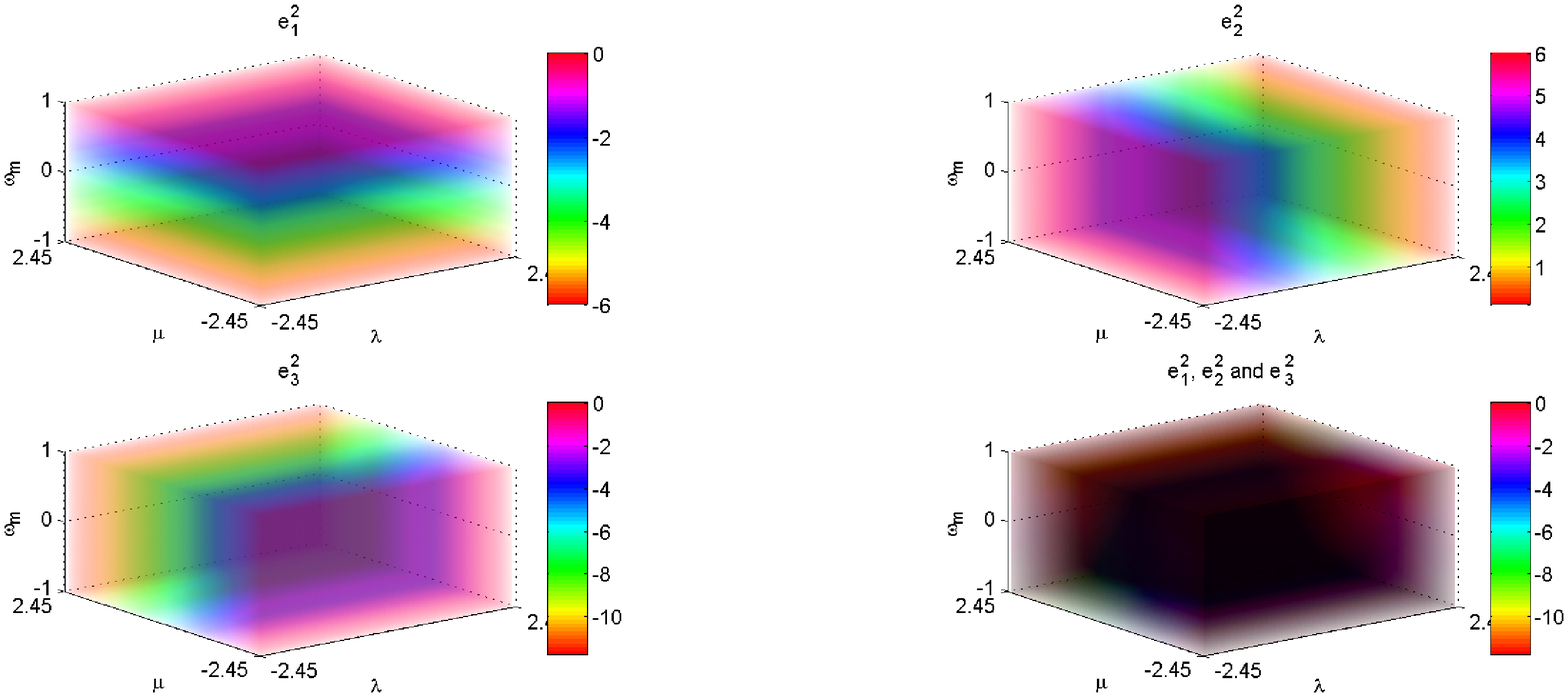}
\caption{$C_2$ }\label{fig102}
\end{center}
\end{figure}

\noindent FIG.\ref{fig102} represents  the stability analysis for $C_2$ for different values of parameters in all the possible cases. In FIG.\ref{fig102}, the stability of $C_2$ in the parameter region $\{-1\le\omega_m\le 1,-\sqrt{6}\le \lambda\le \sqrt{6},-\sqrt{6}\le \mu\le \sqrt{6}\}$ has been considered.

\smallskip

\bigskip

\subsection{Critical Point $C_3$}

\noindent For $C_3$ the $A^3=\left(%
\begin{array}{c}
  1 \\
  0 \\
  \frac{\mu^2}{3}-1 \\
\end{array}%
\right).$

\noindent The Jacobian matrix of the system (\ref{ads})-(\ref{ads2}) at
this point has the characteristic polynomial
\begin{equation}a^3_3s^3+a^3_2s^2+a^3_1s+a^3_0=0\end{equation}
where \begin{equation}a^3_3=1,\end{equation}
\begin{equation}a^3_2=\frac{108\mu-36\omega_m\mu^3-45\mu^2+54\omega_m\mu+9\lambda\mu^2-3\lambda\mu^4+6\omega_m\mu^5+9\mu^3}{2\mu(\mu^2-3)^2},\end{equation} 
 \begin{equation}a^3_1= X3-Y3\end{equation} 
where
\begin{equation}X3=\frac{27\lambda\mu+27\omega_m\mu^2-9\omega_m\mu^4+27\mu^2-9\lambda\omega_m\mu^3+27\lambda\omega_m\mu}{2(\mu^2-3)^2}\end{equation}
and
\begin{equation}Y3= \frac{3(54\mu-27\mu^3+3\mu^5)(6\omega_m+\lambda\mu-2\omega_m\mu^2+\mu^2+6)}{4\mu(\mu^2-3)^3}\end{equation}
and 
\begin{equation}a^3_0=\frac{3(\lambda+\mu)(-3\omega_m\mu^2+9\omega_m+9)(54\mu-27\mu^3+3\mu^5)}{4(\mu^2-3)^4}.\end{equation}

\smallskip

\noindent In this case,

\begin{equation}e^3_1= \frac{3}{\frac{\mu^2}{3}-1}-3\omega_m,\end{equation}
\begin{equation}e^3_2= \frac{3(\mu^2+\lambda\mu)}{2(\mu^2-3)}\end{equation}
and
\begin{equation}e^3_3= \frac{3(-18+9\mu^2-\mu^4)}{2(\mu^2-3)^2}.\end{equation}

\smallskip

\noindent $e^3_1$ is zero if $\mu=\sqrt{3(1+\frac{1}{\omega_m})},$ $e^3_2$ is zero if $\mu=-\lambda,$ where $\lambda<0,$ whereas $e^3_3$ is always non-zero for $C_3$. $e^3_1$ and $e^3_2$ are both zero if $\omega_m=\frac{3}{\lambda^2-3},$ where $\lambda<0.$

\noindent We present the TABLE \ref{tab6} containing the
non-hyperbolic subcases and their stability.

\bigskip
\begin{table}[H]
\centering%
\caption{$C_3$(Cases)}\label{tab6}
\bigskip
\begin{tabular}{|c|c|c|c|}
  \hline
Case & $\omega_m$ & $\lambda$ & $\mu$ \\
  \hline
  a & NA & NA & $\sqrt{3(1+\frac{1}{\omega_m})}$ \\
  \hline
  b &  NA & $\lambda<0$ & $\mu=-\lambda$ \\
  \hline
   c & $\omega_m=\frac{3}{\lambda^2-3}$ & $\lambda<0$ & NA  \\
  \hline
 \end{tabular}
\end{table}

\smallskip

\noindent The center manifolds and the reduced systems are described in the TABLE \ref{tab7}.

\bigskip
\begin{table}[H]
\centering%
\caption{$C_3$(Center Manifolds and Reduced System)}\label{tab7}
\bigskip
\begin{tabular}{|c|c|c|}
  \hline
Case & Center Manifold & Reduced System \\
  \hline
  a & $\bar{\bar{y}}=0,\bar{\bar{\nu}}=C^{3,a}_{2,(2,0,0)}\bar{\bar{x}}^2$ & $\bar{\bar{x}}'= R^{3,a}_{1,(2,0,0)}\bar{\bar{x}}^2$\\
  \hline
  b  & $\bar{\bar{y}}= C^{3,b}_{1,(2,0,0)}\bar{\bar{x}}^2,\bar{\bar{\nu}}=C^{3,b}_{2,(2,0,0)}\bar{\bar{x}}^2$ & $\bar{\bar{x}}'=R^{3,b}_{1,(5,0,0)}\bar{\bar{x}}^5$ \\
  \hline
  c & $\bar{\bar{\nu}}= C^{3,c}_{1,(2,0,0)}\bar{\bar{x}}^2+C^{3,c}_{1,(0,2,0)}\bar{\bar{y}}^2$ & $\bar{\bar{x}}'=R^{3,c}_{1,(1,2,0)} \bar{\bar{x}}\bar{\bar{y}}^2, \bar{\bar{y}}'=R^{3,c}_{2,(0,2,0)}\bar{\bar{y}}^2$  \\
  \hline
 \end{tabular}
\end{table}

\noindent where \begin{equation}C^{3,a}_{2,(2,0,0)}= -\frac{\omega_m(8\omega_m^2+7\omega_m+1)}{4(\omega_m^2-1)},R^{3,a}_{1,(2,0,0)}= -\frac{3}{2}\omega_m^2\end{equation}
and \begin{equation}C^{3,b}_{1,(2,0,0)}=-\frac{\lambda^2(\lambda^2-6)}{3(6\omega_m-2\lambda^2\omega_m+\lambda^2)}, C^{3,b}_{2,(2,0,0)}= -\frac{2\lambda^2(\lambda^2-3)(\omega_m-1)}{3(6\omega_m-2\lambda^2\omega_m+\lambda^2)}\end{equation}
and \begin{equation}R^{3,b}_{1,(5,0,0)}= \frac{3(6\omega_m-2\lambda^2\omega_m-5\lambda^2+6)}{16(\lambda^2-3)}.\end{equation}
\noindent Also,
\begin{equation}C^{3,c}_{1,(2,0,0)}=\frac{2\lambda^2}{3}, C^{3,c}_{1,(0,2,0)}= \frac{12(\lambda^2-3)(2916\lambda^2-972\lambda^4-1215\lambda^6+756\lambda^8-126\lambda^{10}+\lambda^{14})}{16\lambda^4(\lambda^2-3)^5(\lambda^2-6)^2}\end{equation}
and \begin{equation}R^{3,c}_{1,(1,2,0)}=-\frac{27(-5832\lambda^3+6804\lambda^5-2916\lambda^7+540\lambda^9-36\lambda^{11})}{4\lambda^5(\lambda^2-3)^5(\lambda^2-6)^2}, R^{3,c}_{2,(0,2,0)}=-\frac{27}{(\lambda^2-3)^2}.\end{equation}

\noindent For subcase (a), $P^3_a$ is $\left(%
\begin{array}{ccc}
  \frac{\omega_m}{2(\omega_m+1)} & 0 & 0 \\
  0 & 1 & 0 \\
  1 & 0 & 1 \\
\end{array}%
\right).$

\noindent For subcase (b), $P^3_b= \left(%
\begin{array}{ccc}
  0 & \frac{3(6\omega_m-2\lambda^2\omega_m+\lambda^2)}{2\lambda^2(\lambda^2-6)} & 0 \\
  1 & 0 & 0 \\
  0 & 1 & 1 \\
\end{array}%
\right). $

\noindent For subcase (c), $P^3_c= \left(%
\begin{array}{ccc}
  0 & \frac{3(18\lambda-9\lambda^3+\lambda^5)}{2\lambda^3(\lambda^2-3)(\lambda^2-6)} & 0 \\
  1 & 0 & 0 \\
  0 & 1 & 1 \\
\end{array}%
\right).$

\smallskip

\noindent We summarize our results for the stability of the reduced system of $C_3$ in the TABLE \ref{tab8}.

\bigskip
\begin{table}[H]
\centering%
\caption{Summary for the critical point $C_3$(non-hyperbolic cases)}\label{tab8}
\bigskip
\begin{tabular}{|c|c|c|c|c|}
  \hline
Case & $\omega_m$ & $\lambda$ & $\mu$ & Stability (RS)\\
  \hline
  a & NA & NA & $\sqrt{3(1+\frac{1}{\omega_m})}$ & Stable\\
  \hline
  b &  NA & $\lambda<0$ & $\mu=-\lambda$ &  Stable\\
  \hline
   c & $\omega_m=\frac{3}{\lambda^2-3}$ & $\lambda<0$ & NA & Unstable\\
  \hline
 \end{tabular}
\end{table}

\noindent We notice that in the subcase (b) if $\omega_m=\frac{5\lambda^2-6}{6-2\lambda^2},$ then the center manifold reduction  fails as $R^{3,b}_{1,(5,0,0)}=0.$ In this case, some higher degree center manifold reduction is necessary. Case (c) is interesting in a sense that though it is unstable, it does not diverge to infinity, rather converges to some point other than $C_3$ in the neighborhood of its initial position parallel to $\bar{\bar{y}}$ axis.

\begin{figure}[H]
\begin{center}
\includegraphics
[scale=0.45]{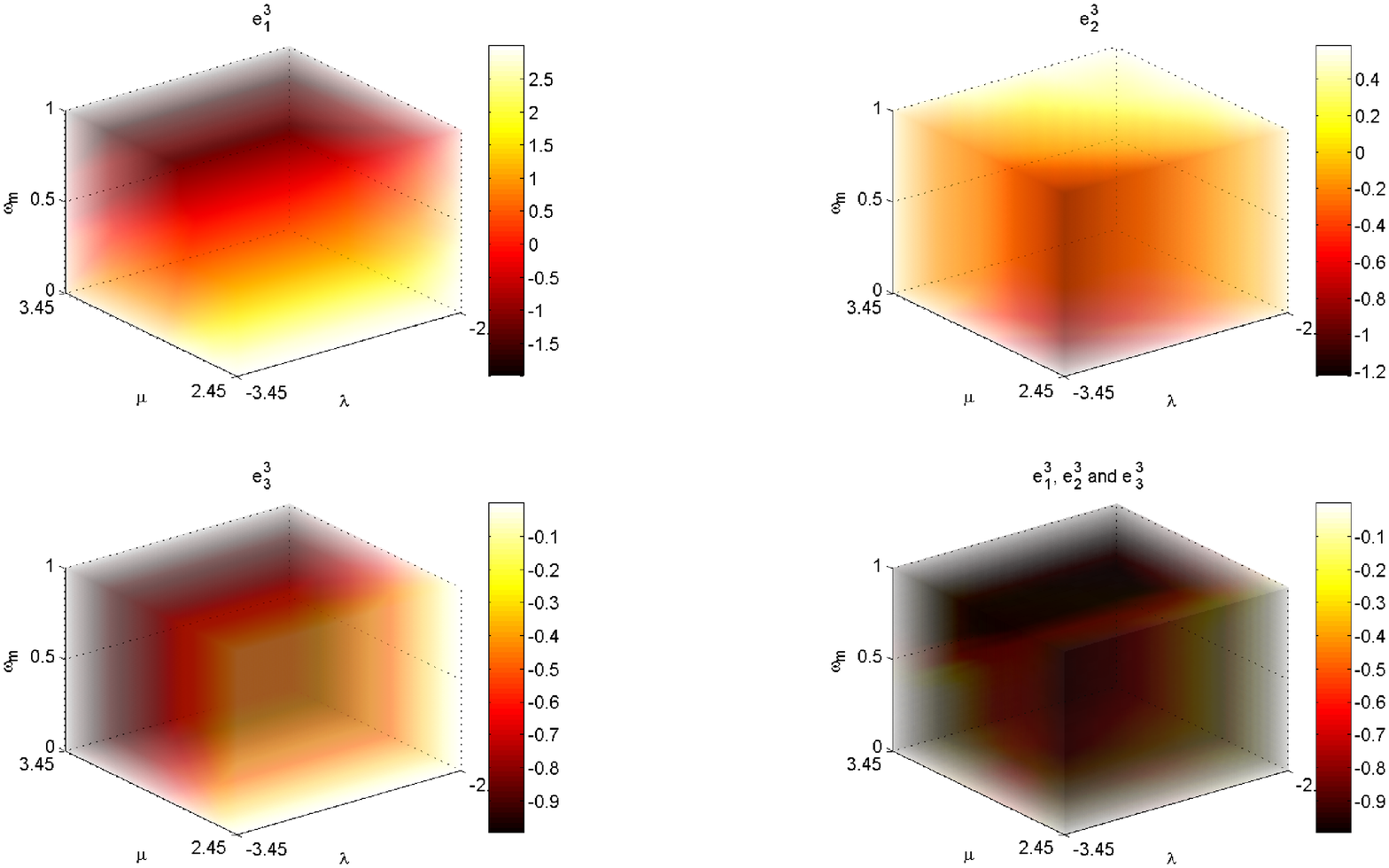}
\caption{$C_3$ }\label{fig103}
\end{center}
\end{figure}

\noindent Lastly we present the FIG.\ref{fig103} to show the stability analysis for $C_3$ for different values of parameters in all the possible cases (both the hyperbolic and the non-hyperbolic). In the FIG.\ref{fig103} the stability of $C_3$ in the region $\{0\le\omega_m\le 1,\sqrt{6}\le \lambda\le \sqrt{6}+1,-\sqrt{6}-1\le \mu\le -\sqrt{6}\}$ of the parameter space has been presented.

\smallskip

\bigskip

\subsection{Critical Point $C_4$}

\noindent For $C_4$ the $A^4=\left(%
\begin{array}{c}
  -1 \\
  0 \\
  \frac{\mu^2}{3}-1 \\
\end{array}%
\right).$

\noindent The Jacobian matrix of the system (\ref{ads})-(\ref{ads2}) has the characteristic polynomial at
this point as 
\begin{equation}a^4_3s^3+a^4_2s^2+a^4_1s+a^4_0=0\end{equation}
where \begin{equation}a^4_3=1,\end{equation}
\begin{equation}a^4_2=\frac{108\mu-36\omega_m\mu^3-45\mu^2+54\omega_m\mu+9\lambda\mu^2-3\lambda\mu^4+6\omega_m\mu^5+9\mu^3}{2\mu(\mu^2-3)^2},\end{equation} 
 \begin{equation}a^4_1= X4-Y4\end{equation} 
where
\begin{equation}X4=\frac{27\lambda\mu+27\omega_m\mu^2-9\omega_m\mu^4+27\mu^2-9\lambda\omega_m\mu^3+27\lambda\omega_m\mu}{2(\mu^2-3)^2}\end{equation}
and
\begin{equation}Y4= \frac{3(54\mu-27\mu^3+3\mu^5)(6\omega_m+\lambda\mu-2\omega_m\mu^2+\mu^2+6)}{4\mu(\mu^2-3)^3}\end{equation}
and 
\begin{equation}a^4_0=\frac{3(\lambda+\mu)(-3\omega_m\mu^2+9\omega_m+9)(54\mu-27\mu^3+3\mu^5)}{4(\mu^2-3)^4}.\end{equation}

\smallskip

\noindent The eigen values are

\begin{equation}e^4_1= \frac{3}{\frac{\mu^2}{3}-1}-3\omega_m,\end{equation}
\begin{equation}e^4_2= \frac{3(\mu^2+\lambda\mu)}{2(\mu^2-3)}\end{equation}
and
\begin{equation}e^4_3= \frac{3(-18+9\mu^2-\mu^4)}{2(\mu^2-3)^2}.\end{equation}

\smallskip

\noindent $e^4_1$ is zero if $\mu=-\sqrt{3(1+\frac{1}{\omega_m})},$ $e^4_2$ is zero if $\mu=-\lambda,$ where $\lambda>0,$ whereas $e^4_3$ is always non-zero for $C_4$. $e^4_1$ and $e^4_2$ are both zero if $\omega_m=\frac{3}{\lambda^2-3},$ where $\lambda>0.$

\noindent We present the TABLE \ref{tab9} containing the
non-hyperbolic subcases and their stability.

\bigskip
\begin{table}[H]
\centering%
\caption{$C_4$(Cases)}\label{tab9}
\bigskip
\begin{tabular}{|c|c|c|c|}
  \hline
Case & $\omega_m$ & $\lambda$ & $\mu$ \\
  \hline
  a & NA & NA & $-\sqrt{3(1+\frac{1}{\omega_m})}$ \\
  \hline
  b &  NA & $\lambda>0$ & $\mu=-\lambda$ \\
  \hline
   c & $\omega_m=\frac{3}{\lambda^2-3}$ & $\lambda>0$ & NA  \\
  \hline
 \end{tabular}
\end{table}

\smallskip

\noindent The center manifolds and the reduced systems are described in the TABLE \ref{tab10}.

\bigskip
\begin{table}[H]
\centering%
\caption{$C_4$(Center Manifolds and Reduced System)}\label{tab10}
\bigskip
\begin{tabular}{|c|c|c|}
  \hline
Case & Center Manifold & Reduced System \\
  \hline
  a & $\bar{\bar{y}}=0,\bar{\bar{\nu}}=C^{4,a}_{2,(2,0,0)}\bar{\bar{x}}^2$ & $\bar{\bar{x}}'= R^{4,a}_{1,(2,0,0)}\bar{\bar{x}}^2$\\
  \hline
  b  & $\bar{\bar{y}}= C^{4,b}_{1,(2,0,0)}\bar{\bar{x}}^2,\bar{\bar{\nu}}=C^{4,b}_{2,(2,0,0)}\bar{\bar{x}}^2$ & $\bar{\bar{x}}'=R^{4,b}_{1,(5,0,0)}\bar{\bar{x}}^5$ \\
  \hline
  c & $\bar{\bar{\nu}}= C^{4,c}_{1,(2,0,0)}\bar{\bar{x}}^2+C^{4,c}_{1,(0,2,0)}\bar{\bar{y}}^2$ & $\bar{\bar{x}}'=R^{4,c}_{1,(1,2,0)} \bar{\bar{x}}\bar{\bar{y}}^2, \bar{\bar{y}}'=R^{4,c}_{2,(0,2,0)}\bar{\bar{y}}^2$  \\
  \hline
 \end{tabular}
\end{table}

\noindent where \begin{equation}C^{4,a}_{2,(2,0,0)}= -\frac{\omega_m(8\omega_m^2+7\omega_m+1)}{4(\omega_m^2-1)},R^{4,a}_{1,(2,0,0)}= -3\omega_m^2\end{equation}
and \begin{equation}C^{4,b}_{1,(2,0,0)}=-\frac{\lambda^2(\lambda^2-6)}{3(6\omega_m-2\lambda^2\omega_m+\lambda^2)}, C^{4,b}_{2,(2,0,0)}= -\frac{2\lambda^2(\lambda^2-3)(\omega_m-1)}{3(6\omega_m-2\lambda^2\omega_m+\lambda^2)}\end{equation}
and \begin{equation}R^{4,b}_{1,(5,0,0)}= \frac{3(6\omega_m-2\lambda^2\omega_m-5\lambda^2+6)}{16(\lambda^2-3)}.\end{equation}
\noindent Also,
\begin{equation}C^{4,c}_{1,(2,0,0)}=\frac{2\lambda^2}{3}, C^{4,c}_{1,(0,2,0)}= \frac{3(\lambda^4+15\lambda^2+18)}{4\lambda^2(\lambda^4-9\lambda^2+18)}\end{equation}
and \begin{equation}R^{4,c}_{1,(1,2,0)}= \frac{243}{\lambda^2(\lambda^2-3)^2(\lambda^2-6)}, R^{4,c}_{2,(0,2,0)}=-\frac{27}{(\lambda^2-3)^2}.\end{equation}

\noindent For subcase (a), $P^4_a$ is $\left(%
\begin{array}{ccc}
  -\frac{\omega_m}{2(\omega_m+1)} & 0 & 0 \\
  0 & 1 & 0 \\
  1 & 0 & 1 \\
\end{array}%
\right).$

\noindent For subcase (b), $P^4_b= \left(%
\begin{array}{ccc}
  0 & -\frac{3(6\omega_m-2\lambda^2\omega_m+\lambda^2)}{2\lambda^2(\lambda^2-6)} & 0 \\
  1 & 0 & 0 \\
  0 & 1 & 1 \\
\end{array}%
\right). $

\noindent For subcase (c), $P^4_c= \left(%
\begin{array}{ccc}
  0 & -\frac{3}{2\lambda^2} & 0 \\
  1 & 0 & 0 \\
  0 & 1 & 1 \\
\end{array}%
\right).$

\smallskip

\noindent In the TABLE \ref{tab11}, we summarize our results for the stability of the reduced system of $C_4.$

\bigskip
\begin{table}[H]
\centering%
\caption{Summary for the critical point $C_4$(non-hyperbolic cases)}\label{tab11}
\bigskip
\begin{tabular}{|c|c|c|c|c|}
  \hline
Case & $\omega_m$ & $\lambda$ & $\mu$ & Stability (RS)\\
  \hline
  a & NA & NA & $-\sqrt{3(1+\frac{1}{\omega_m})}$ & Stable\\
  \hline
  b &  NA & $\lambda>0$ & $\mu=-\lambda$ &  Stable\\
  \hline
   c & $\omega_m=\frac{3}{\lambda^2-3}$ & $\lambda>0$ & NA & Unstable\\
  \hline
 \end{tabular}
\end{table}

\noindent Again here we notice that in the subcase (b) if $\omega_m=\frac{5\lambda^2-6}{6-2\lambda^2},$ then the center manifold reduction  fails as $R^{4,b}_{1,(5,0,0)}=0.$ So, some higher degree center manifold reduction is necessary. Case (c) is again interesting because although it is unstable, it does not diverge to infinity, rather converges to some point other than $C_4$ in the neighborhood of its initial position parallel to $\bar{\bar{y}}$ axis.

\begin{figure}[H]
\begin{center}
\includegraphics
[scale=0.45]{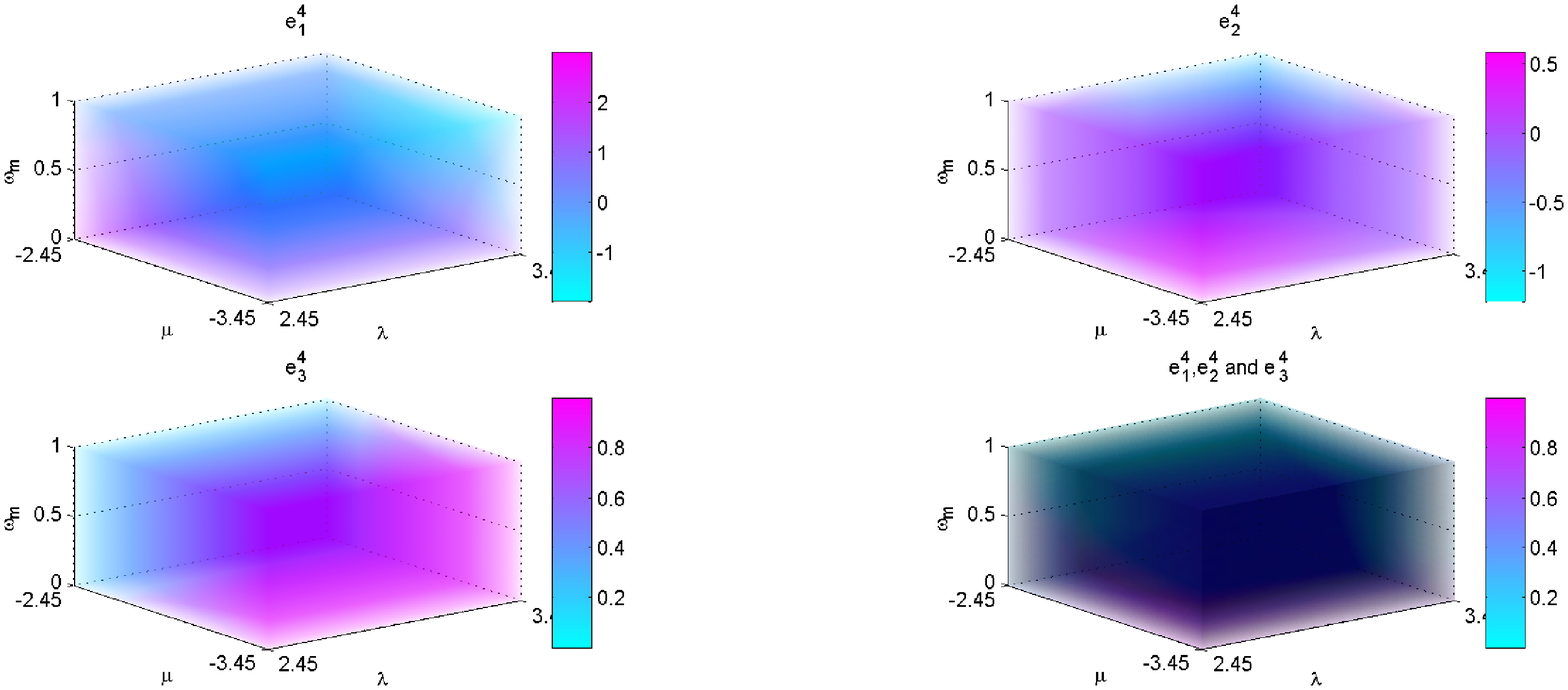}
\caption{$C_4$ }\label{fig104}
\end{center}
\end{figure}

\noindent Next we present the FIG.\ref{fig104} to show the stability analysis for $C_4$ for different values of parameters in all the possible cases to end this subsection. We note that in the FIG.\ref{fig104}, we took $\{0\le\omega_m\le 1,\sqrt{6}\le \lambda\le \sqrt{6}+1,-\sqrt{6}-1\le \mu\le -\sqrt{6}\}$ as our parameter space.

\smallskip

\bigskip

\subsection{Critical Point $C_5$}

\noindent For $C_5$ the $A^5=\left(%
\begin{array}{c}
  \frac{\sqrt{3(1+\frac{1}{\omega_m})}}{\mu}\\
  0 \\
  \frac{1}{\omega_m} \\
\end{array}%
\right).$

\noindent The Jacobian matrix of the system (\ref{ads})-(\ref{ads2}) at
$C_5$ has the characteristic polynomial as :
\begin{equation}a^5_3s^3+a^5_2s^2+a^5_1s+a^5_0=0\end{equation}
where \begin{equation}a^5_3=1,\end{equation}
\begin{equation}a^5_2= -[\frac{3\mu^2(\omega_m-1)}{2\mu^2}+\frac{3(\lambda+\mu+\omega_m\lambda+\omega_m\mu)}{2\mu}],\end{equation} 
 \begin{equation}a^3_1= X5+Y5\end{equation} 
where
\begin{equation}X5= \frac{(\omega_m+1)(27-27\omega_m^2-9\omega_m\mu^2+9\omega_m^2\mu^2)}{2\mu^2}\end{equation}
and
\begin{equation}Y5= \frac{9(\omega_m-1)(\lambda+\mu+\lambda\omega_m+\omega_m\mu)}{4\mu}\end{equation}
and 
\begin{equation}a^5_0= -\frac{3(\lambda+\mu+\omega_m\lambda+\omega_m\mu)(\omega_m+1)(27-27\omega_m^2-9\omega_m\mu^2+9\omega_m^2\mu^2)}{4\mu^3}.\end{equation}

\smallskip

\noindent Here,

\begin{equation}e^5_1= \frac{3(\omega_m\mu-\mu+\sqrt{(1-\omega_m)(8\omega_m^2\mu^2+7\omega_m\mu^2-24\omega_m^2+\mu^2-48\omega_m-24)})}{4\mu},\end{equation}
\begin{equation}e^5_2=\frac{3(\omega_m\mu-\mu-\sqrt{(1-\omega_m)(8\omega_m^2\mu^2+7\omega_m\mu^2-24\omega_m^2+\mu^2-48\omega_m-24)})}{4\mu}\end{equation}
and
\begin{equation}e^5_3= \frac{3(1+\omega_m)(\lambda+\mu)}{2\mu}.\end{equation}

\smallskip

\noindent $e^5_1$ and $e^5_2$ are both zero if $\mu=\sqrt{3(1+\frac{1}{\omega_m})}$ or if $\mu=-\sqrt{3(1+\frac{1}{\omega_m})}.$ But then $C_5$ coinsides with $C_3$ in the first case and with $C_4$ in the second case. $e^5_3$ is zero if $\mu=-\lambda$ where $\lambda\neq 0.$ So this is the only new case.

\noindent For this case, the center manifold and the reduced system are described in the TABLE \ref{tab12}.

\smallskip
\begin{table}[H]
\centering%
\caption{$C_5$(Center Manifolds and Reduced System)}\label{tab12}
\bigskip
\begin{tabular}{|c|c|c|}
  \hline
Case & Center Manifold & Reduced System \\
  \hline
  a & $\bar{\bar{y}}=C^{5,a}_{1,(2,0,0)}\bar{\bar{x}}^2,\bar{\bar{\nu}}=C^{5,a}_{2,(2,0,0)}\bar{\bar{x}}^2$ & $\bar{\bar{x}}'= R^{5,a}_{1,(5,0,0)}\bar{\bar{x}}^5$\\
  \hline
 \end{tabular}
\end{table}

\noindent where \begin{equation}C^{5,a}_{1,(2,0,0)}= -\frac{\lambda^2(\lambda+3\omega_m\lambda-4\lambda\omega_m^2+Z5)}{6\omega_mZ5}, C^{5,a}_{2,(2,0,0)}= \frac{\lambda^2(\lambda+3\omega_m\lambda-4\lambda\omega_m^2-Z5)}{6\omega_mZ5}\end{equation}
and \begin{equation}R^{5,a}_{1,(5,0,0)}= -\frac{(4\lambda^4\omega_m+24\lambda^2\omega_m^2+24\lambda^2\omega_m+\lambda^4)}{48(1+\omega_m)}. \end{equation}

\smallskip

\noindent Here $Z5=\sqrt{(1-\omega_m)(8\omega_m^2\lambda^2+7\omega_m\lambda^2-24\omega_m^2+\lambda^2-48\omega_m-24)}.$

\smallskip

\noindent For this subcase, $P^5_a$ is $\left(%
\begin{array}{ccc}
  0 &  \frac{\sqrt{3\omega_m(1+\omega_m)}(\lambda-\lambda\omega_m+Z5)}{4\lambda^2(\omega_m^2-1)}& \frac{\sqrt{3\omega_m(1+\omega_m)}(\lambda-\lambda\omega_m-Z5)}{4\lambda^2(\omega_m^2-1)} \\
  1 & 0& 0 \\
  0 & 1 & 1 \\
\end{array}%
\right).$

\smallskip

\noindent We summarize our results for the stability of the reduced system of $C_5$ in the TABLE \ref{tab13}.

\smallskip
\begin{table}[H]
\centering%
\caption{Summary for the critical point $C_5$(non-hyperbolic cases)}\label{tab13}
\bigskip
\begin{tabular}{|c|c|c|c|c|}
  \hline
Case & $\omega_m$ & $\lambda$ & $\mu$ & Stability (RS)\\
  \hline
  a & NA & $\lambda\neq 0$ & $-\lambda$ & Stable\\
  \hline
 \end{tabular}
\end{table}

\smallskip

\begin{figure}[H]
\begin{center}
\includegraphics
[scale=0.45]{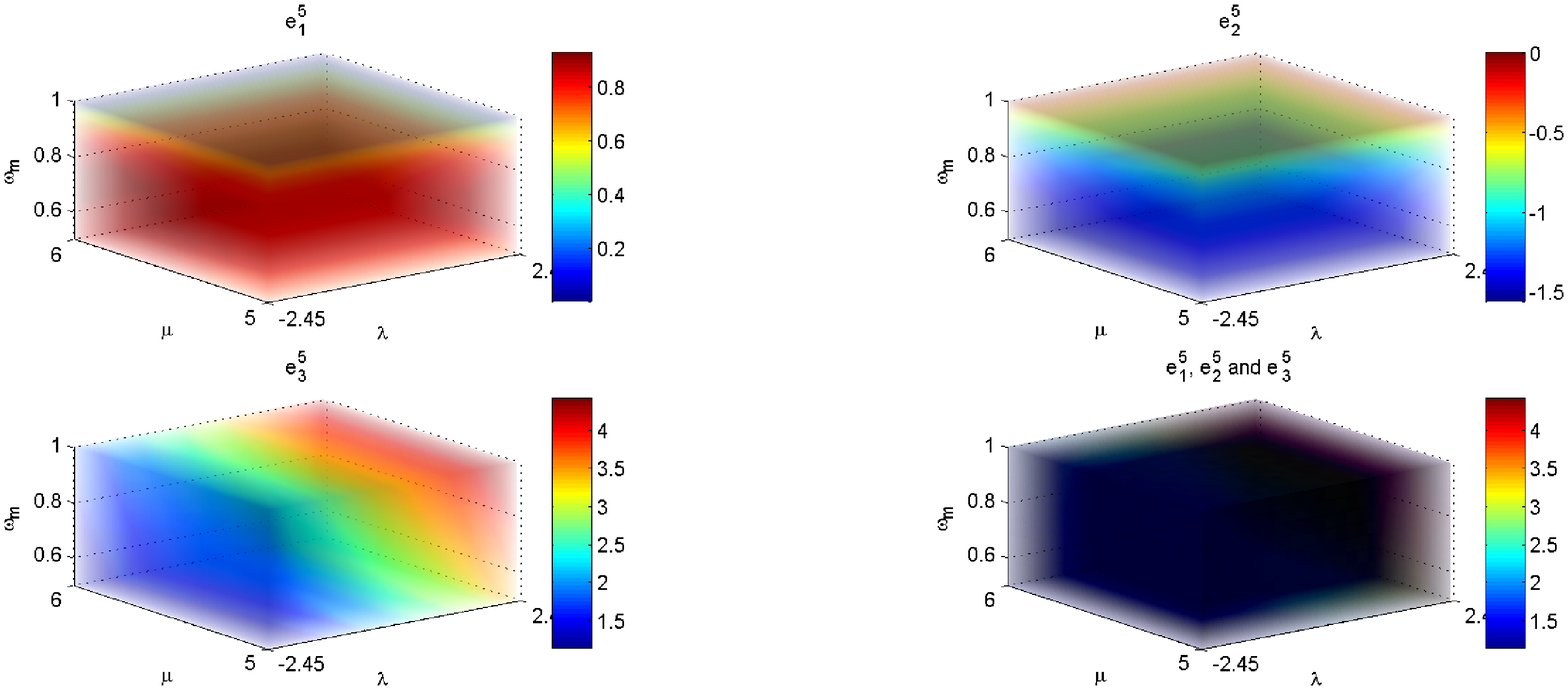}
\caption{$C_5$ }\label{fig105}
\end{center}
\end{figure}

\noindent  Now we present FIG.\ref{fig105} showing the stability analysis for $C_5$ for different values of parameters to conclude this subsection. In the FIG.\ref{fig105}, the values of the parameters have been restricted in the region $\{0\le\omega_m\le 1,-\sqrt{6}\le \lambda\le \sqrt{6},5\le \mu\le 6\}.$

\smallskip

\bigskip

\subsection{Critical Point $C_6$}

\noindent For $C_6$ the $A^6=\left(%
\begin{array}{c}
  -\frac{\sqrt{3(1+\frac{1}{\omega_m})}}{\mu}\\
  0 \\
  \frac{1}{\omega_m} \\
\end{array}%
\right).$

\noindent The Jacobian matrix of the system (\ref{ads})-(\ref{ads2}) at
$C_6$ has the characteristic polynomial as the following :
\begin{equation}a^6_3s^3+a^6_2s^2+a^6_1s+a^6_0=0\end{equation}
where \begin{equation}a^6_3=1,\end{equation}
\begin{equation}a^6_2= \frac{3(4\mu\omega_m^2+2\mu+\omega_m\lambda+\lambda)}{2\mu},\end{equation} 
 \begin{equation}a^6_1= \frac{9(1+\omega_m)(-6\omega_m^2\mu^2+4\lambda\omega_m^2\mu+6\omega_m^2+\omega_m\mu^2+\lambda\omega_m\mu-3\mu^2+3\lambda\mu-6)}{4\mu^2}\end{equation} 
and 
\begin{equation}a^6_0= \frac{27(\omega_m^2-1)(\lambda-\mu)(\omega_m+1)(-\omega_m\mu^2+3\omega_m+3)}{4\mu^3}.\end{equation}

\smallskip

\noindent Then,

\begin{equation}e^6_1= -\frac{3(3\mu+\omega_m\mu+4\omega_m^2\mu+Y6)}{4\mu},\end{equation}
\begin{equation}e^6_2=-\frac{3(3\mu+\omega_m\mu+4\omega_m^2\mu-Y6)}{4\mu}\end{equation}
and
\begin{equation}e^6_3= -\frac{3(1+\omega_m)(\lambda-\mu)}{2\mu}.\end{equation}

\smallskip

\noindent Where $Y6= \sqrt{(16\omega_m^4\mu^2+16\omega_m^3\mu^2-24\omega_m^3+25\omega_m^2\mu^2-24\omega_m^2-2\omega_m\mu^2+24\omega_m+9\mu^2+24)}.$

\smallskip

\noindent Here again, $e^6_1$ and $e^6_2$ are both zero if $\mu=\sqrt{3(1+\frac{1}{\omega_m})}$ or if $\mu=-\sqrt{3(1+\frac{1}{\omega_m})}.$ But then $C_6$ coinsides with $C_4$ in the first case and with $C_3$ in the second case. $e^6_3$ is zero if $\mu=\lambda$ where $\lambda\neq 0.$ But this case also identical with subcase (a) of $C_5.$ So no new case arise with $C_6.$

\smallskip

\begin{figure}[H]
\begin{center}
\includegraphics
[scale=0.45]{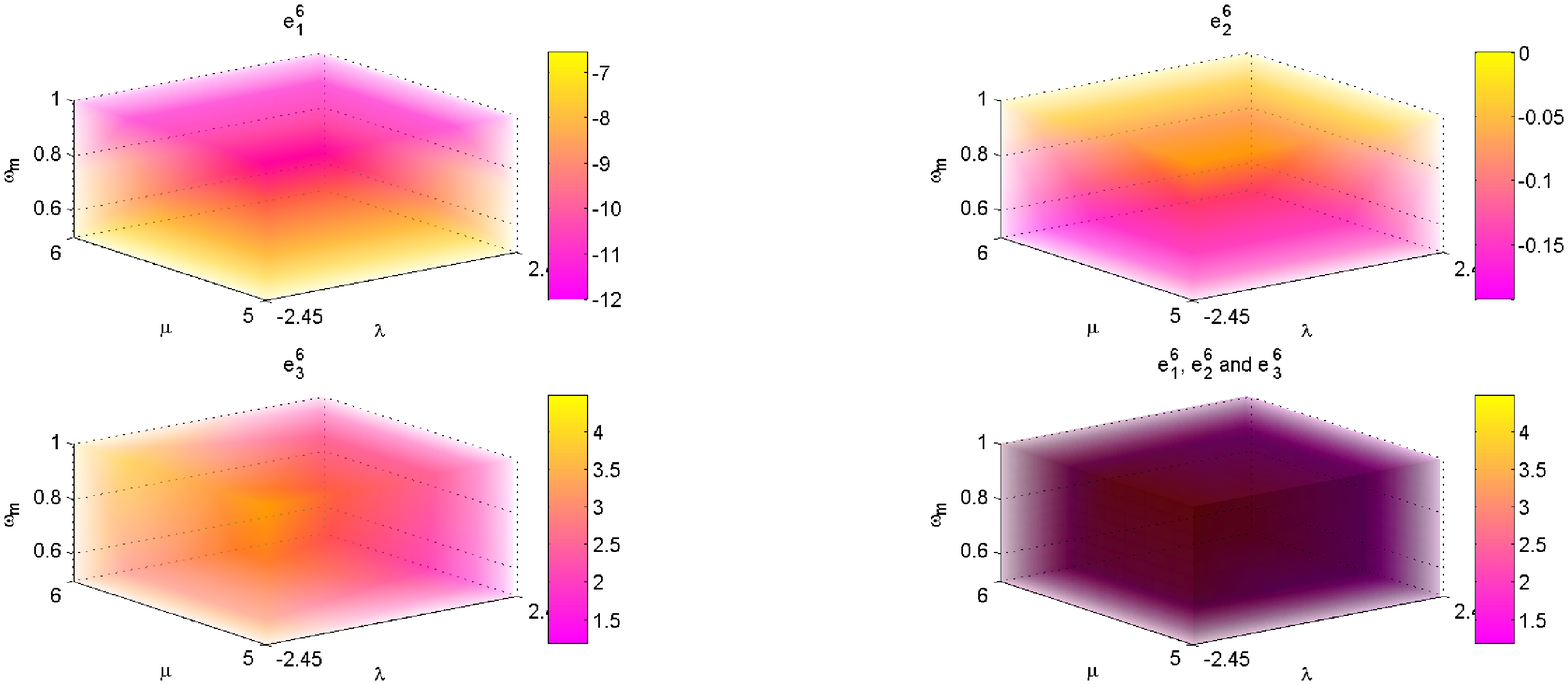}
\caption{$C_6$ }\label{fig106}
\end{center}
\end{figure}

\noindent We present FIG.\ref{fig106} showing the stability analysis for $C_6$ for different values of parameters for the possible cases and conclude this subsection. It is to be noted that the parameter space in the FIG.\ref{fig106} is $\{0\le\omega_m\le 1,-\sqrt{6}\le \lambda\le \sqrt{6},5\le \mu\le 6\}.$

\smallskip

\bigskip

\subsection{Critical Point $C_7$}

\noindent For $C_7$ the $A^7=\left(%
\begin{array}{c}
  -\frac{\lambda}{\sqrt{6}}\\
  \sqrt{\frac{6-\lambda^2}{6}} \\
  1 \\
\end{array}%
\right).$

\noindent The Jacobian matrix of the system (\ref{ads})-(\ref{ads2}) has the characteristic polynomial at
this point as 
\begin{equation}a^7_3s^3+a^7_2s^2+a^7_1s+a^7_0=0\end{equation}
where \begin{equation}a^7_3=1,\end{equation}
\begin{equation}a^7_2= -\frac{\lambda^2}{2}+\mu\lambda+3\omega_m+6,\end{equation} 
 \begin{equation}a^7_1= 9\omega_m+6\lambda\mu+\frac{3\lambda^2\omega_m}{2}-\frac{3\lambda^3\mu}{2}+\frac{3\lambda^2}{2}-\lambda^4+3\lambda\omega_m\mu+9\end{equation} 
and 
\begin{equation}a^7_0= -\frac{\lambda(\lambda^2-6)(\lambda+\mu)(-\lambda^2+3\omega_m+3)}{2}.\end{equation}

\smallskip

\noindent The eigen values are

\begin{equation}e^7_1= \frac{\lambda^2}{2}-3,\end{equation}
\begin{equation}e^7_2= -\lambda(\lambda+\mu)\end{equation}
and
\begin{equation}e^7_3= \lambda^2-3\omega_m-3.\end{equation}

\smallskip

\noindent $e^7_1$ is never zero for $C_7$.  $e^7_2$ is zero if $\mu=-\lambda,$ where $e^7_3$ is zero for $\omega_m=\frac{\lambda^2}{3}-1$. $e^7_2$ and $e^7_3$ are both zero if $\mu=-\lambda$ and $\omega_m=\frac{\lambda^2}{3}-1$ hold together.

\noindent We present the TABLE \ref{tab14} containing the
non-hyperbolic subcases and their stability.

\bigskip
\begin{table}[H]
\centering%
\caption{$C_7$(Cases)}\label{tab14}
\bigskip
\begin{tabular}{|c|c|c|c|}
  \hline
Case & $\omega_m$ & $\lambda$ & $\mu$ \\
  \hline
  a & NA & $|\lambda|<\sqrt{6},\lambda\neq 0$ & $-\lambda$ \\
  \hline
  b &  $\frac{\lambda^2}{3}-1$ & $|\lambda|<\sqrt{6},\lambda\neq 0$ & NA \\
  \hline
  c &  $\frac{\lambda^2}{3}-1$ & $|\lambda|<\sqrt{6},\lambda\neq 0$ &  $-\lambda$  \\
  \hline
 \end{tabular}
\end{table}

\smallskip

\noindent The center manifolds and the reduced systems are described in the TABLE \ref{tab15}.

\bigskip
\begin{table}[H]
\centering%
\caption{$C_7$(Center Manifolds and Reduced System)}\label{tab15}
\bigskip
\begin{tabular}{|c|c|c|}
  \hline
Case & Center Manifold & Reduced System \\
  \hline
  a & $\bar{\bar{y}}=C^{7,a}_{1,(2,0,0)}\bar{\bar{x}}^2,\bar{\bar{\nu}}=C^{7,a}_{2,(2,0,0)}\bar{\bar{x}}^2$ & $\bar{\bar{x}}'= R^{7,a}_{1,(4,0,0)}\bar{\bar{x}}^4$\\
  \hline
  b  & $\bar{\bar{y}}= C^{7,b}_{1,(2,0,0)}\bar{\bar{x}}^2,\bar{\bar{\nu}}= 0$ & $\bar{\bar{x}}'=R^{7,b}_{1,(2,0,0)}\bar{\bar{x}}^2$ \\
  \hline
  c & $\bar{\bar{\nu}}= C^{7,c}_{1,(2,0,0)}\bar{\bar{x}}^2+C^{7,c}_{1,(1,1,0)}\bar{\bar{x}}\bar{\bar{y}}+C^{7,c}_{1,(0,2,0)}\bar{\bar{y}}^2$ & $\bar{\bar{x}}'=R^{7,c}_{1,(2,0,0)} \bar{\bar{x}}^2+R^{7,c}_{1,(1,1,0)}\bar{\bar{x}}\bar{\bar{y}}+R^{7,c}_{1,(0,2,0)}\bar{\bar{y}}^2$\\

     & &  $\bar{\bar{y}}'=R^{7,c}_{2,(2,1,0)}\bar{\bar{x}}^2\bar{\bar{y}}+R^{7,c}_{2,(1,2,0)}\bar{\bar{x}}\bar{\bar{y}}^2+ R^{7,c}_{2,(0,3,0)}\bar{\bar{y}}^3$\\
\hline  
 \end{tabular}
\end{table}

\noindent where \begin{equation}C^{7,a}_{1,(2,0,0)}= -\frac{\sqrt{6}\lambda^2(36\omega_m-4\lambda^2\omega_m-8\lambda^2+\lambda^4)}{(6-\lambda^2)^{\frac{3}{2}}(-64\lambda^2+384\omega_m)}, C^{7,a}_{2,(2,0,0)}=  -\frac{\sqrt{6}\lambda^2\omega_m}{32\sqrt{(6-\lambda^2)}(-\lambda^2+6\omega_m)}\end{equation}
and \begin{equation}R^{7,a}_{1,(4,0,0)}= \frac{(18\lambda^4-288\lambda^2+648)}{128(\lambda^2-6)^2}.\end{equation}
Then \begin{equation}C^{7,b}_{1,(2,0,0)}= -\frac{\sqrt{6}\lambda^2(2\lambda^4-15\lambda^2+18)}{2(\lambda^2-3)^2(6-\lambda^2)^\frac{3}{2}}, R^{7,b}_{1,(2,0,0)}= \frac{\sqrt{6}\lambda^2(2\lambda^2-12)}{2(\lambda^2-3)\sqrt{(6-\lambda^2)}}.\end{equation}
\noindent Also,
\begin{equation}C^{7,c}_{1,(2,0,0)}=  -\frac{\sqrt{6}\lambda^2(2\lambda^4-15\lambda^2+18)}{2(\lambda^2-3)^2(6-\lambda^2)^\frac{3}{2}},C^{7,c}_{1,(1,1,0)}= \frac{(-2\lambda^4+3\lambda^2)}{4(\lambda^2-3)^2}\end{equation}
and \begin{equation}C^{7,c}_{1,(0,2,0)}= \frac{\sqrt{6}\lambda^2(-5\lambda^6+9\lambda^4+81\lambda^2-162)}{192(\lambda^2-3)^2(6-\lambda^2)^\frac{3}{2}}.\end{equation}
The reduced system coefficients for subcase (c) are :
\begin{equation}R^{7,c}_{1,(2,0,0)}=  \frac{\sqrt{6}\lambda^2(2\lambda^2-12)}{2(\lambda^2-3)\sqrt{(6-\lambda^2)}}, R^{7,c}_{1,(1,1,0)}= -\frac{\lambda^2(\lambda^2-6)}{2(\lambda^2-3)}\end{equation}
and
\begin{equation}R^{7,c}_{1,(0,2,0)}= \frac{\sqrt{6}\lambda^4(6\lambda^4-60\lambda^2+144)}{192\sqrt{(6-\lambda^2)}(\lambda^4-9\lambda^2+18)}\end{equation}
and
\begin{equation} R^{7,c}_{2,(2,1,0)}= \frac{36\lambda^2}{(\lambda^2-3)^2}, R^{7,c}_{2,(1,2,0)}= -\frac{3\lambda^2\sqrt{(36-6\lambda^2)}}{(\lambda^2-3)^2}\end{equation}
and
\begin{equation}R^{7,c}_{2,(0,3,0)}= \frac{9\lambda^4(\lambda^2-4)}{8(\lambda^2-3)^2(\lambda^2-6)}.\end{equation}

\noindent For subcase (a), $P^7_a$ is $\left(%
\begin{array}{ccc}
  \frac{\sqrt{6}\lambda}{24} & \frac{\sqrt{(6-\lambda^2)}}{\lambda} & -\frac{\lambda(\omega_m-1)}{\omega_m\sqrt{(6-\lambda^2)}} \\
  \frac{\sqrt{6}\lambda^2}{24\sqrt{(6-\lambda^2)}} & 1 & 1 \\
  1 & 0 & 0 \\
\end{array}%
\right).$

\noindent For subcase (b), $P^7_b= \left(%
\begin{array}{ccc}
  -\frac{\lambda(6-\lambda^2)^{\frac{3}{2}}}{(\lambda^4-9\lambda^2+18)} & \frac{\sqrt{(6-\lambda^2)}}{\lambda} & \frac{\sqrt{6}\lambda(\lambda^2-6)}{24(3\lambda^2+2\mu\lambda-6)} \\
  1 & 1 &  -\frac{\lambda^2\sqrt{(36-6\lambda^2)}}{24(3\lambda^2+2\mu\lambda-6)}\\
  0 & 0 & 1 \\
\end{array}%
\right). $

\noindent For subcase (c), $P^7_c= \left(%
\begin{array}{ccc}
   -\frac{\lambda(6-\lambda^2)^{\frac{3}{2}}}{(\lambda^4-9\lambda^2+18)}  & -\frac{\sqrt{6}\lambda}{8(\lambda^2-3)} & \frac{\sqrt{(6-\lambda^2)}}{\lambda} \\
  1 & 0 & 1 \\
  0 & 1 & 0 \\
\end{array}%
\right).$

\smallskip

\noindent In the TABLE \ref{tab16}, we summarize our results for the stability of the reduced system of $C_7.$

\bigskip
\begin{table}[H]
\centering%
\caption{Summary for $C_7$(non-hyperbolic cases)}\label{tab16}
\bigskip
\begin{tabular}{|c|c|c|c|c|}
  \hline
Case & $\omega_m$ & $\lambda$ & $\mu$ & Stability (RS)\\
  \hline
  a & NA & $|\lambda|<\sqrt{6},\lambda\neq 0$ & $-\lambda$ & Stable\\
  \hline
  b &  $\frac{\lambda^2}{3}-1$ & $|\lambda|<\sqrt{6},\lambda\neq 0$ & NA  & Stable\\
  \hline
  c &  $\frac{\lambda^2}{3}-1$ & $|\lambda|<\sqrt{6},\lambda\neq 0$ &  $-\lambda$  & Unstable\\
  \hline
 \end{tabular}
\end{table}

\noindent Where the coupled nonlinear system of ODEs representing case (c) is chaotic in nature and hence is unstable.
We also observe that when $\lambda=\sqrt{7}-1,\mu=1-\sqrt{7}$ and $\omega_m\neq \frac{5-2\sqrt{7}}{3}$ or $\lambda=-\sqrt{7}+1,\mu=-1+\sqrt{7}$ and $\omega_m\neq \frac{5-2\sqrt{7}}{3}$ in the subcase (a), then $R^{7,a}_{1,(4,0,0)}=0.$ Therefore we can not decide the stability. Some higher degree center manifold reduction seems to be nesessary. 

\begin{figure}[H]
\begin{center}
\includegraphics
[scale=0.45]{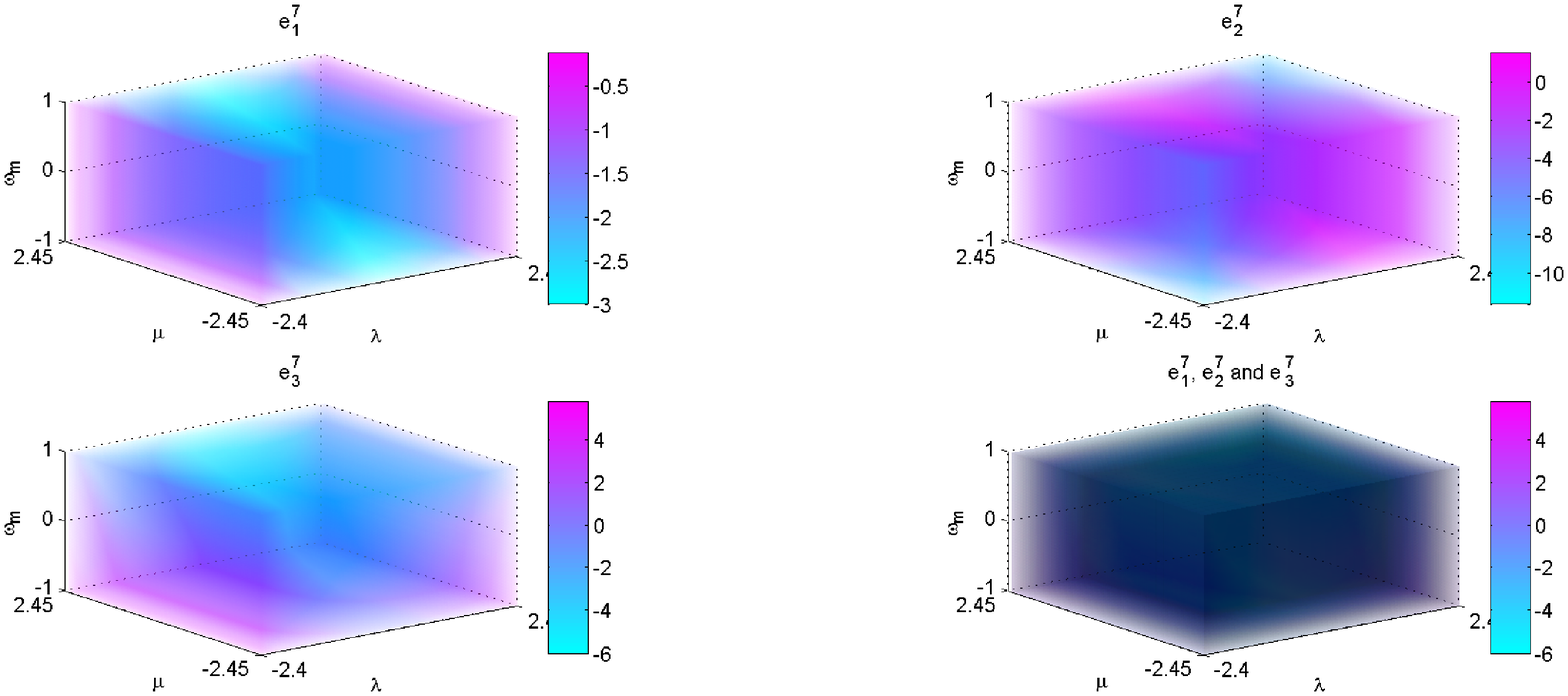}
\caption{$C_7$ }\label{fig107}
\end{center}
\end{figure}

\noindent Next we present FIG.\ref{fig107} showing the stability analysis for $C_7$ and end this subsection. The parameter space in the FIG.\ref{fig107} is the set $\{-1\le\omega_m\le 1,-\sqrt{6}\le \lambda\le \sqrt{6},-\sqrt{6}\le \mu\le \sqrt{6}\}.$

\smallskip

\bigskip

\subsection{Critical Point $C_8$}

\noindent For this case, $A^8=\left(%
\begin{array}{c}
  -\frac{\lambda}{\sqrt{6}}\\
  -\sqrt{\frac{6-\lambda^2}{6}} \\
  1 \\
\end{array}%
\right).$

\noindent The Jacobian matrix of the system (\ref{ads})-(\ref{ads2}) has the characteristic polynomial

\begin{equation}a^8_3s^3+a^8_2s^2+a^8_1s+a^8_0=0\end{equation}
where \begin{equation}a^8_3=1,\end{equation}
\begin{equation}a^8_2= -\frac{\lambda^2}{2}+\mu\lambda+3\omega_m+6,\end{equation} 
 \begin{equation}a^8_1= 9\omega_m+6\lambda\mu+\frac{3\lambda^2\omega_m}{2}-\frac{3\lambda^3\mu}{2}+\frac{3\lambda^2}{2}-\lambda^4+3\lambda\omega_m\mu+9\end{equation} 
and 
\begin{equation}a^8_0= -\frac{\lambda(\lambda^2-6)(\lambda+\mu)(-\lambda^2+3\omega_m+3)}{2}.\end{equation}

\smallskip

\noindent The eigen values are

\begin{equation}e^8_1= \frac{\lambda^2}{2}-3,\end{equation}
\begin{equation}e^8_2= -\lambda(\lambda+\mu)\end{equation}
and
\begin{equation}e^8_3= \lambda^2-3\omega_m-3.\end{equation}

\smallskip

\noindent $e^8_1$ is never zero.  $e^8_2$ is zero if $\mu=-\lambda,$ where $e^8_3$ is zero for $\omega_m=\frac{\lambda^2}{3}-1$. $e^8_2$ and $e^8_3$ are both zero if $\mu=-\lambda$ and $\omega_m=\frac{\lambda^2}{3}-1$ hold together.

\noindent We present the TABLE \ref{tab17} containing various subcases and their stability.

\bigskip
\begin{table}[H]
\centering%
\caption{$C_8$(Cases)}\label{tab17}
\bigskip
\begin{tabular}{|c|c|c|c|}
  \hline
Case & $\omega_m$ & $\lambda$ & $\mu$ \\
  \hline
  a & NA & $|\lambda|<\sqrt{6},\lambda\neq 0$ & $-\lambda$ \\
  \hline
  b &  $\frac{\lambda^2}{3}-1$ & $|\lambda|<\sqrt{6},\lambda\neq 0$ & NA \\
  \hline
  c &  $\frac{\lambda^2}{3}-1$ & $|\lambda|<\sqrt{6},\lambda\neq 0$ &  $-\lambda$  \\
  \hline
 \end{tabular}
\end{table}

\smallskip

\noindent The center manifolds and the reduced systems are described in the TABLE \ref{tab18}.

\bigskip
\begin{table}[H]
\centering%
\caption{$C_8$(Center Manifolds and Reduced System)}\label{tab18}
\bigskip
\begin{tabular}{|c|c|c|}
  \hline
Case & Center Manifold & Reduced System \\
  \hline
  a & $\bar{\bar{y}}=C^{8,a}_{1,(2,0,0)}\bar{\bar{x}}^2,\bar{\bar{\nu}}=C^{8,a}_{2,(2,0,0)}\bar{\bar{x}}^2$ & $\bar{\bar{x}}'= R^{8,a}_{1,(4,0,0)}\bar{\bar{x}}^4$\\
  \hline
  b  & $\bar{\bar{y}}= C^{8,b}_{1,(2,0,0)}\bar{\bar{x}}^2,\bar{\bar{\nu}}= 0$ & $\bar{\bar{x}}'=R^{8,b}_{1,(2,0,0)}\bar{\bar{x}}^2$ \\
  \hline
  c & $\bar{\bar{\nu}}= C^{8,c}_{1,(2,0,0)}\bar{\bar{x}}^2+C^{8,c}_{1,(1,1,0)}\bar{\bar{x}}\bar{\bar{y}}+C^{8,c}_{1,(0,2,0)}\bar{\bar{y}}^2$ & $\bar{\bar{x}}'=R^{8,c}_{1,(2,0,0)} \bar{\bar{x}}^2+R^{8,c}_{1,(1,1,0)}\bar{\bar{x}}\bar{\bar{y}}+R^{8,c}_{1,(0,2,0)}\bar{\bar{y}}^2$\\

     & &  $\bar{\bar{y}}'=R^{8,c}_{2,(2,1,0)}\bar{\bar{x}}^2\bar{\bar{y}}+R^{8,c}_{2,(1,2,0)}\bar{\bar{x}}\bar{\bar{y}}^2+ R^{8,c}_{2,(0,3,0)}\bar{\bar{y}}^3$\\
\hline  
 \end{tabular}
\end{table}

\noindent where \begin{equation}C^{8,a}_{1,(2,0,0)}= \frac{\sqrt{6}\lambda^2(36\omega_m-4\lambda^2\omega_m-8\lambda^2+\lambda^4)}{(6-\lambda^2)^{\frac{3}{2}}(-64\lambda^2+384\omega_m)}, C^{8,a}_{2,(2,0,0)}=  \frac{\sqrt{6}\lambda^2\omega_m}{32\sqrt{(6-\lambda^2)}(-\lambda^2+6\omega_m)}\end{equation}
and \begin{equation}R^{8,a}_{1,(4,0,0)}= \frac{(18\lambda^4-288\lambda^2+648)}{128(\lambda^2-6)^2}.\end{equation}
Then \begin{equation}C^{8,b}_{1,(2,0,0)}= \frac{\sqrt{6}\lambda^2(2\lambda^4-15\lambda^2+18)}{2(\lambda^2-3)^2(6-\lambda^2)^\frac{3}{2}}, R^{8,b}_{1,(2,0,0)}= -\frac{\sqrt{6}\lambda^2(2\lambda^2-12)}{2(\lambda^2-3)\sqrt{(6-\lambda^2)}}.\end{equation}
\noindent Also,
\begin{equation}C^{8,c}_{1,(2,0,0)}=  \frac{\sqrt{6}\lambda^2(2\lambda^4-15\lambda^2+18)}{2(\lambda^2-3)^2(6-\lambda^2)^\frac{3}{2}},C^{8,c}_{1,(1,1,0)}= \frac{(-2\lambda^4+3\lambda^2)}{4(\lambda^2-3)^2}\end{equation}
and \begin{equation}C^{8,c}_{1,(0,2,0)}=-\frac{\sqrt{6}\lambda^2(-5\lambda^6+9\lambda^4+81\lambda^2-162)}{192(\lambda^2-3)^2(6-\lambda^2)^\frac{3}{2}}.\end{equation}
The reduced system coefficients for subcase (c) are :
\begin{equation}R^{8,c}_{1,(2,0,0)}=  -\frac{\sqrt{6}\lambda^2(2\lambda^2-12)}{2(\lambda^2-3)\sqrt{(6-\lambda^2)}}, R^{8,c}_{1,(1,1,0)}= -\frac{\lambda^2(\lambda^2-6)}{2(\lambda^2-3)}\end{equation}
and
\begin{equation}R^{8,c}_{1,(0,2,0)}= -\frac{\sqrt{6}\lambda^4(6\lambda^4-60\lambda^2+144)}{192\sqrt{(6-\lambda^2)}(\lambda^4-9\lambda^2+18)}\end{equation}
and
\begin{equation} R^{8,c}_{2,(2,1,0)}= \frac{36\lambda^2}{(\lambda^2-3)^2}, R^{8,c}_{2,(1,2,0)}= \frac{3\lambda^2\sqrt{(36-6\lambda^2)}}{(\lambda^2-3)^2}\end{equation}
and
\begin{equation}R^{8,c}_{2,(0,3,0)}= \frac{9\lambda^4(\lambda^2-4)}{8(\lambda^2-3)^2(\lambda^2-6)}.\end{equation}

\noindent For subcase (a), $P^8_a$ is $\left(%
\begin{array}{ccc}
  \frac{\sqrt{6}\lambda}{24} & -\frac{\sqrt{(6-\lambda^2)}}{\lambda} & \frac{\lambda(\omega_m-1)}{\omega_m\sqrt{(6-\lambda^2)}} \\
  -\frac{\sqrt{6}\lambda^2}{24\sqrt{(6-\lambda^2)}} & 1 & 1 \\
  1 & 0 & 0 \\
\end{array}%
\right).$

\noindent For subcase (b), $P^8_b= \left(%
\begin{array}{ccc}
  \frac{\lambda(6-\lambda^2)^{\frac{3}{2}}}{(\lambda^4-9\lambda^2+18)} & -\frac{\sqrt{(6-\lambda^2)}}{\lambda} & \frac{\sqrt{6}\lambda(\lambda^2-6)}{24(3\lambda^2+2\mu\lambda-6)} \\
  1 & 1 &  \frac{\lambda^2\sqrt{(36-6\lambda^2)}}{24(3\lambda^2+2\mu\lambda-6)}\\
  0 & 0 & 1 \\
\end{array}%
\right). $

\noindent For subcase (c), $P^8_c= \left(%
\begin{array}{ccc}
   \frac{\lambda(6-\lambda^2)^{\frac{3}{2}}}{(\lambda^4-9\lambda^2+18)}  & -\frac{\sqrt{6}\lambda}{8(\lambda^2-3)} & -\frac{\sqrt{(6-\lambda^2)}}{\lambda} \\
  1 & 0 & 1 \\
  0 & 1 & 0 \\
\end{array}%
\right).$

\smallskip

\noindent In the TABLE \ref{tab19}, we summarize our results for the stability of the reduced system of $C_8.$

\bigskip
\begin{table}[H]
\centering%
\caption{Summary for $C_8$(non-hyperbolic scenario)}\label{tab19}
\bigskip
\begin{tabular}{|c|c|c|c|c|}
  \hline
Case & $\omega_m$ & $\lambda$ & $\mu$ & Stability (RS)\\
  \hline
  a & NA & $|\lambda|<\sqrt{6},\lambda\neq 0$ & $-\lambda$ & Stable\\
  \hline
  b &  $\frac{\lambda^2}{3}-1$ & $|\lambda|<\sqrt{6},\lambda\neq 0$ & NA  & Stable\\
  \hline
  c &  $\frac{\lambda^2}{3}-1$ & $|\lambda|<\sqrt{6},\lambda\neq 0$ &  $-\lambda$  & Unstable\\
  \hline
 \end{tabular}
\end{table}

\noindent Again we see that for case (c) the dynamical system governing the reduced system of $C_8$ is chaotic in nature. It is also observed that in the subcase (a), when $\lambda=\sqrt{7}-1,\mu=1-\sqrt{7}$ and $\omega_m\neq \frac{5-2\sqrt{7}}{3}$ or $\lambda=-\sqrt{7}+1,\mu=-1+\sqrt{7}$ and $\omega_m\neq \frac{5-2\sqrt{7}}{3}$ , then $R^{8,a}_{1,(4,0,0)}=0.$ Therefore we can't decide the stability, some higher degree center manifold reduction seems nesessary for this particular case. 

\begin{figure}[H]
\begin{center}
\includegraphics
[scale=0.45]{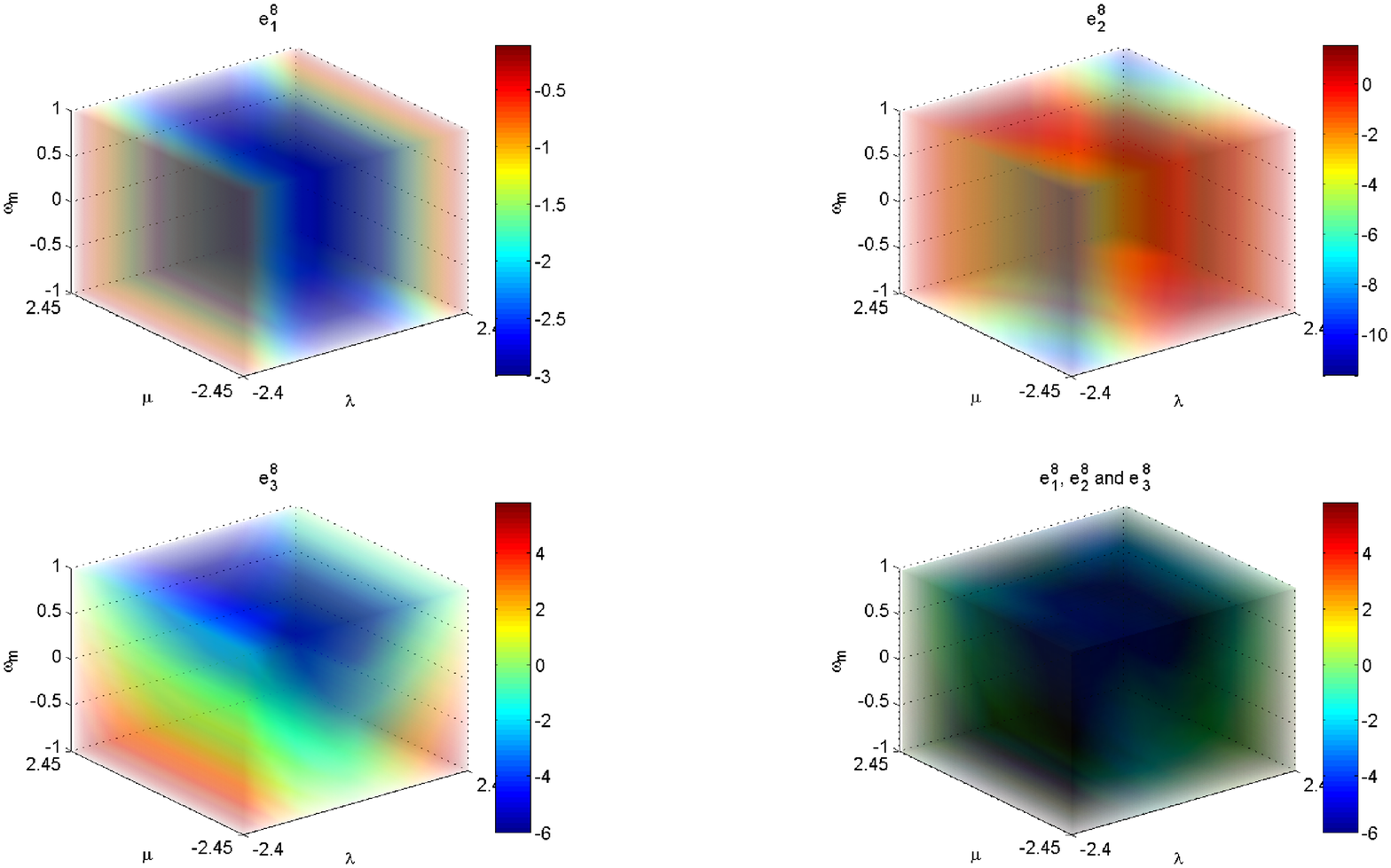}
\caption{$C_8$ }\label{fig108}
\end{center}
\end{figure}

\noindent We end this subsection after presenting FIG.\ref{fig108} showing the stability analysis for $C_8$ for different values of parameters. The parameter space considered in FIG.\ref{fig108} is $\{-1\le\omega_m\le 1,-\sqrt{6}\le \lambda\le \sqrt{6},-\sqrt{6}\le \mu\le \sqrt{6}\}.$

\smallskip

\bigskip

\bigskip

\subsection{Critical Point $C_9$}

\noindent For this case, $A^9=\left(%
\begin{array}{c}
 -\frac{\sqrt{6}(1+\omega_m)}{2\lambda}\\
  \frac{\sqrt{6(1-\omega_m^2)}}{2\lambda} \\
  1 \\
\end{array}%
\right).$

\noindent The jacobian matrix of the system (\ref{ads})-(\ref{ads2}) has the following characteristic polynomial

\begin{equation}a^9_3s^3+a^9_2s^2+a^9_1s+a^9_0=0\end{equation}
where \begin{equation}a^9_3=1,\end{equation}
\begin{equation}a^9_2= \frac{(9\lambda+6\mu+3\lambda\omega_m+6\omega_m\mu)}{2\lambda},\end{equation} 
 \begin{equation}a^9_1= \frac{9(\omega_m^2-1)(-2\lambda^2-\mu\lambda+3\omega_m+3)}{2\lambda^2}\end{equation} 
and 
\begin{equation}a^9_0= \frac{27(\lambda+\mu)(\omega_m-1)(\omega_m+1)^2(-\lambda^2+3\omega_m+3)}{2\lambda^3}.\end{equation}

\smallskip

\noindent The eigen values are

\begin{equation}e^9_1= \frac{3(\lambda\omega_m-\lambda+X9)}{4\lambda},\end{equation}
\begin{equation}e^9_2= -\frac{3(\lambda+\mu)(\omega_m+1)}{\lambda}\end{equation}
and
\begin{equation}e^9_3= \frac{3(\lambda\omega_m-\lambda-X9)}{4\lambda},\end{equation}
where
\begin{equation}X9= \sqrt{(1-\omega_m)(-9\lambda^2\omega_m-7\lambda^2+24\omega_m^2+48\omega_m+24)}.\end{equation}

\smallskip

\noindent $e^9_1$ and $e^9_3$ are both zero if $\omega_m=1$ or $\omega_m=\frac{\lambda^2}{3}-1.$ But in the later subcase $C_9$ becomes identical with $C_7$ or $C_8$ depending on whether $\lambda>0$ or $\lambda<0$ respectively. $e^9_2$ is zero if $\mu=-\lambda.$ Also in the subcase where $\omega_m=1$ and $\mu=-\lambda$ both holds, $C_9$ becomes identical with $C_5$ or $C_6$ depending on the sign of $\lambda.$ Hence $\omega_m=1$ and $\mu=-\lambda$ separately are the only interesting subcases. In the first subcase the first and the third eigenvalues are both zero, whereas in the second subcase the second eigenvalue is zero.

\noindent We present the TABLE \ref{tab20} containing the
non-hyperbolic subcases and their stability.

\bigskip
\begin{table}[H]
\centering%
\caption{$C_9$(Cases)}\label{tab20}
\bigskip
\begin{tabular}{|c|c|c|c|}
  \hline
Case & $\omega_m$ & $\lambda$ & $\mu$ \\
  \hline
  a & $1$ & $\lambda\neq 0$ & NA\\
  \hline
  b &  $-1<\omega_m\leq 1$ & $\lambda\neq 0$ & $-\lambda$\\
  \hline
 \end{tabular}
\end{table}

\smallskip

\noindent The center manifolds and the reduced systems are described by the TABLE \ref{tab21}.

\bigskip
\begin{table}[H]
\centering%
\caption{$C_9$(Center Manifolds and Reduced System)}\label{tab21}
\bigskip
\begin{tabular}{|c|c|c|}
  \hline
Case & Center Manifold & Reduced System \\
  \hline
  a & $\bar{\bar{\nu}}=0$ & $\bar{\bar{x}}'= R^{9,a}_{1,(0,2,0)}\bar{\bar{y}}^2, \bar{\bar{y}}'= R^{9,a}_{2,(1,1,0)}\bar{\bar{x}}\bar{\bar{y}}$\\
  \hline
  b  & $\bar{\bar{y}}= C^{9,b}_{1,(2,0,0)}\bar{\bar{x}}^2,\bar{\bar{\nu}}=  C^{9,b}_{2,(2,0,0)}\bar{\bar{x}}^2$ & $\bar{\bar{x}}'=R^{9,b}_{1,(4,0,0)}\bar{\bar{x}}^4$ \\
  \hline
 \end{tabular}
\end{table}

\noindent where \begin{equation}R^{9,a}_{1,(0,2,0)}= -\frac{\sqrt{6}(\lambda^2-6)}{2\lambda}, R^{9,a}_{2,(1,1,0)}=  \frac{\sqrt{6}\lambda}{2}\end{equation}
and 
\begin{equation}C^{9,b}_{1,(2,0,0)}= \frac{\sqrt{6}(1+\omega_m)^2(Y9+Z9X9)}{128\lambda^2(1-\omega_m^2)^{\frac{3}{2}}X9},C^{9,b}_{2,(2,0,0)}= - \frac{\sqrt{6}(1+\omega_m)^2(Y9-Z9X9)}{128\lambda^2(1-\omega_m^2)^{\frac{3}{2}}X9}\end{equation}
\noindent and
\begin{equation}R^{9,b}_{1,(4,0,0)}=  \frac{18\omega_m^2-60\omega_m+90}{128(1-\omega_m)^2},\end{equation}
where
\begin{equation}Y9=24\omega_m+\lambda^2\omega_m+5\lambda^2+24\omega_m^2-24\omega_m^3-24\omega_m^4-21\lambda^2\omega_m^2+15\lambda^2\omega_m^3\end{equation}
and
\begin{equation}Z9= -5\lambda-5\lambda\omega_m^2+6\lambda\omega_m.\end{equation}

\noindent For subcase (a), $P^9_a$ is $\left(%
\begin{array}{ccc}
 1 & 0 & \frac{\sqrt{6}(\lambda^2-6)}{4\lambda^2(\lambda+\mu)}  \\
 0 & 1 & 0 \\
  0 & 0 & 1 \\
\end{array}%
\right).$

\noindent For subcase (b), $P^9_b= \left(%
\begin{array}{ccc}
 -\frac{3\sqrt{6}(1+\omega_m)}{8\lambda} & M9+N9 & -M9+N9 \\
  \frac{\sqrt{6(1-\omega_m^2)}(3\omega_m-1)}{8\lambda(\omega_m-1)} & 1 & 1\\
  1 & 0 & 0 \\
\end{array}%
\right), $

\noindent where

\begin{equation}M9= \frac{\lambda(\lambda\omega_m-\lambda+X9)}{2\sqrt{(1-\omega_m^2)}(\lambda^2+3\omega_m^2-3)}\end{equation}
and
\begin{equation}N9= -\frac{3(\omega_m-1)(\omega_m+1)^2}{\sqrt{(1-\omega_m^2)}(\lambda^2+3\omega_m^2-3)}.\end{equation}

\smallskip

\noindent In the TABLE \ref{tab22}, we summarize our results for the stability of the reduced system of $C_9.$

\bigskip
\begin{table}[H]
\centering%
\caption{Summary for $C_9$(non-hyperbolic scenario)}\label{tab22}
\bigskip
\begin{tabular}{|c|c|c|c|c|}
  \hline
Case & $\omega_m$ & $\lambda$ & $\mu$ & Stability(RS)\\
  \hline
  a & $1$ & $\lambda\neq 0$ & NA & Unstable\\
  \hline
  b &  $-1<\omega_m\leq 1$ & $\lambda\neq 0$ & $-\lambda$ & Stable\\
  \hline
 \end{tabular}
\end{table}

\begin{figure}[H]
\begin{center}
\includegraphics
[scale=0.45]{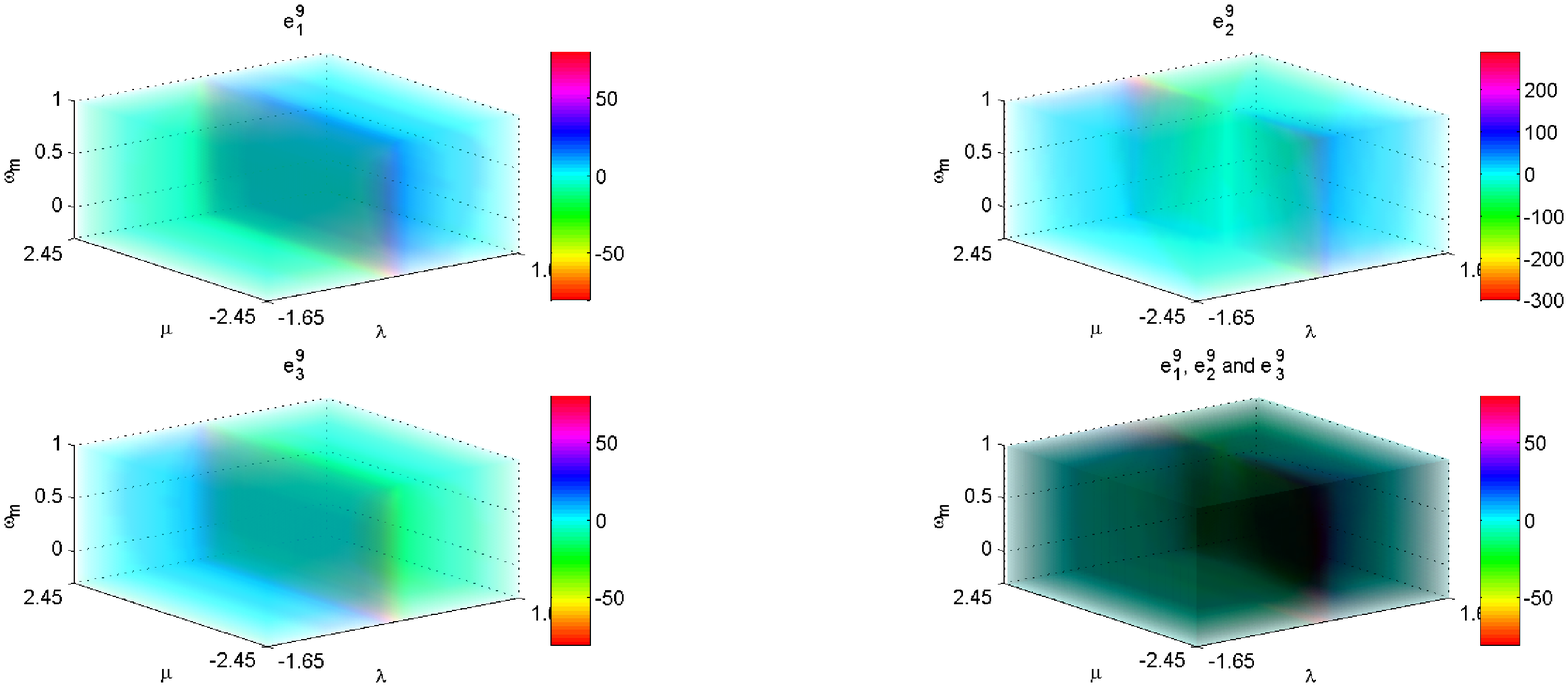}
\caption{$C_9$ }\label{fig109}
\end{center}
\end{figure}

\noindent For the subcase (a) the reduced system is unstable/chaotic in nature.
\noindent Next we present FIG. \ref{fig109} showing the stability analysis for $C_9$  to end this subsection. The parameter space considered here is $\{0\le\omega_m\le 1,-1.65\le \lambda\le 1.65,-\sqrt{6}\le \mu\le \sqrt{6}\}$.

\smallskip

\bigskip

\subsection{Critical Point $C_{10}$}

\noindent For this case, $A^{10}=\left(%
\begin{array}{c}
 -\frac{\sqrt{6}(1+\omega_m)}{2\lambda}\\
  -\frac{\sqrt{6(1-\omega_m^2)}}{2\lambda} \\
  1 \\
\end{array}%
\right).$

\noindent The jacobian matrix of the system (\ref{ads})-(\ref{ads2}) has the characteristic polynomial

\begin{equation}a^{10}_3s^3+a^{10}_2s^2+a^{10}_1s+a^{10}_0=0\end{equation}
where \begin{equation}a^{10}_3=1,\end{equation}
\begin{equation}a^{10}_2= \frac{(9\lambda+6\mu+3\lambda\omega_m+6\omega_m\mu)}{2\lambda},\end{equation} 
 \begin{equation}a^{10}_1= \frac{9(\omega_m^2-1)(-2\lambda^2-\mu\lambda+3\omega_m+3)}{2\lambda^2}\end{equation} 
and 
\begin{equation}a^{10}_0= -\frac{27(\lambda+\mu)(\omega_m-1)(\omega_m+1)^2(-\lambda^2+3\omega_m+3)}{2\lambda^3}.\end{equation}

\smallskip

\noindent Then,

\begin{equation}e^{10}_1= \frac{3(\lambda\omega_m-\lambda+X9)}{4\lambda},\end{equation}
\begin{equation}e^{10}_2= -\frac{3(\lambda+\mu)(\omega_m+1)}{\lambda}\end{equation}
and
\begin{equation}e^{10}_3= \frac{3(\lambda\omega_m-\lambda-X9)}{4\lambda}.\end{equation}

\smallskip

\noindent $e^{10}_1$ and $e^{10}_3$ are both zero if $\omega_m=1$ or $\omega_m=\frac{\lambda^2}{3}-1.$ But in the first subcase $C_{10}$ reduces to $C_9$ and in the later subcase $C_{10}$ becomes identical with $C_7$ or $C_8$ depending on whether $\lambda<0$ or $\lambda>0.$ $e^{10}_2$ is zero if $\mu=-\lambda.$ Hence $\mu=-\lambda$ is the only interesting subcase.

\noindent We present the TABLE \ref{tab23} containing the
non-hyperbolic subcases and their stability.

\bigskip
\begin{table}[H]
\centering%
\caption{$C_{10}$(Cases)}\label{tab23}
\bigskip
\begin{tabular}{|c|c|c|c|}
  \hline
Case & $\omega_m$ & $\lambda$ & $\mu$ \\
  \hline
  a &  $-1<\omega_m\leq 1$ & $\lambda\neq 0$ & $-\lambda$\\
  \hline
 \end{tabular}
\end{table}

\smallskip

\noindent The center manifolds and the reduced systems are described by the TABLE \ref{tab24}.

\bigskip
\begin{table}[H]
\centering%
\caption{$C_{10}$(Center Manifolds and Reduced System)}\label{tab24}
\bigskip
\begin{tabular}{|c|c|c|}
  \hline
Case & Center Manifold & Reduced System \\
  \hline
  a  & $\bar{\bar{y}}= C^{{10},a}_{1,(2,0,0)}\bar{\bar{x}}^2,\bar{\bar{\nu}}=  C^{{10},a}_{2,(2,0,0)}\bar{\bar{x}}^2$ & $\bar{\bar{x}}'=R^{{10},a}_{1,(4,0,0)}\bar{\bar{x}}^4$ \\
  \hline
 \end{tabular}
\end{table}

\noindent where
\begin{equation}C^{{10},a}_{1,(2,0,0)}= -\frac{\sqrt{6}(1+\omega_m)^2(Y9+Z9X9)}{128\lambda^2(1-\omega_m^2)^{\frac{3}{2}}X9},C^{{10},a}_{2,(2,0,0)}= \frac{\sqrt{6}(1+\omega_m)^2(Y9-Z9X9)}{128\lambda^2(1-\omega_m^2)^{\frac{3}{2}}X9}\end{equation}
\noindent and
\begin{equation}R^{{10},a}_{1,(4,0,0)}=  \frac{18\omega_m^2-60\omega_m+90}{128(1-\omega_m)^2},\end{equation}
where $X9,Y9$ and $Z9$ are as defined for $C_9.$

\smallskip

\noindent For subcase (a), $P^{10}_a= \left(%
\begin{array}{ccc}
 -\frac{3\sqrt{6}(1+\omega_m)}{8\lambda} & -M9-N9 & M9-N9 \\
  -\frac{\sqrt{6(1-\omega_m^2)}(3\omega_m-1)}{8\lambda(\omega_m-1)} & 1 & 1\\
  1 & 0 & 0 \\
\end{array}%
\right), $

\noindent where $M9$ and $N9$ are as defined in the previous subsection.

\smallskip

\noindent In the TABLE \ref{tab25}, we summarize our results for the stability of the reduced system of $C_{10}.$

\bigskip
\begin{table}[H]
\centering%
\caption{Summary for $C_{10}$(non-hyperbolic scenario)}\label{tab25}
\bigskip
\begin{tabular}{|c|c|c|c|c|}
  \hline
Case & $\omega_m$ & $\lambda$ & $\mu$ & Stability(RS)\\
  \hline
  a &  $-1<\omega_m\leq 1$ & $\lambda\neq 0$ & $-\lambda$ & Stable\\
  \hline
 \end{tabular}
\end{table}

\begin{figure}[H]
\begin{center}
\includegraphics
[scale=0.45]{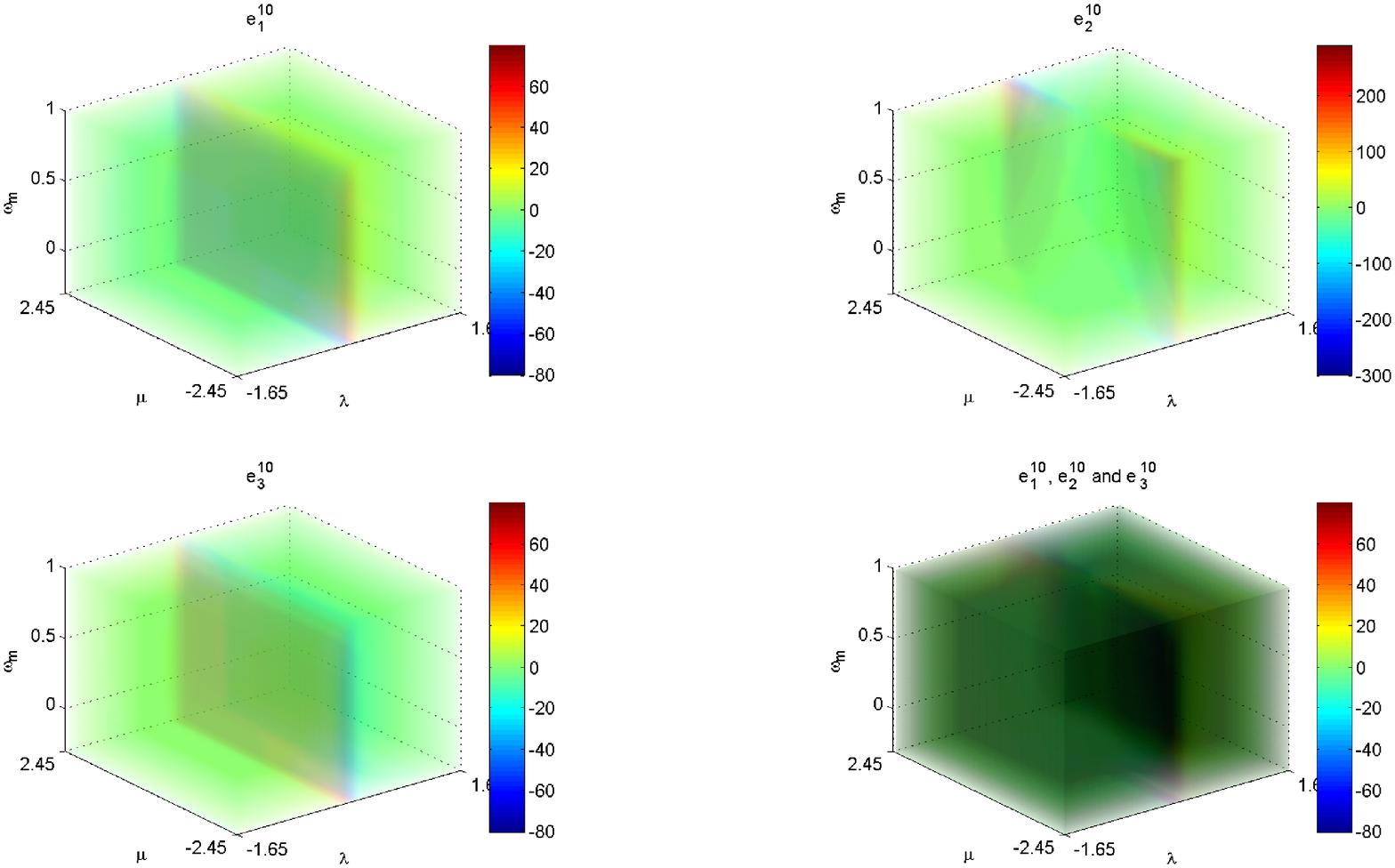}
\caption{$C_{10}$ }\label{fig110}
\end{center}
\end{figure}

\noindent Lastly we present FIG.\ref{fig110} showing the stability analysis for $C_{10}$  to end our analysis of the critical points.  The parameter space considered in FIG.\ref{fig110} is $\{0\le\omega_m\le 1,-1.65\le \lambda\le 1.65,-\sqrt{6}\le \mu\le \sqrt{6}\}$.

\smallskip

\bigskip

\begin{figure}[H]
\begin{center}
\includegraphics
[scale=0.75]{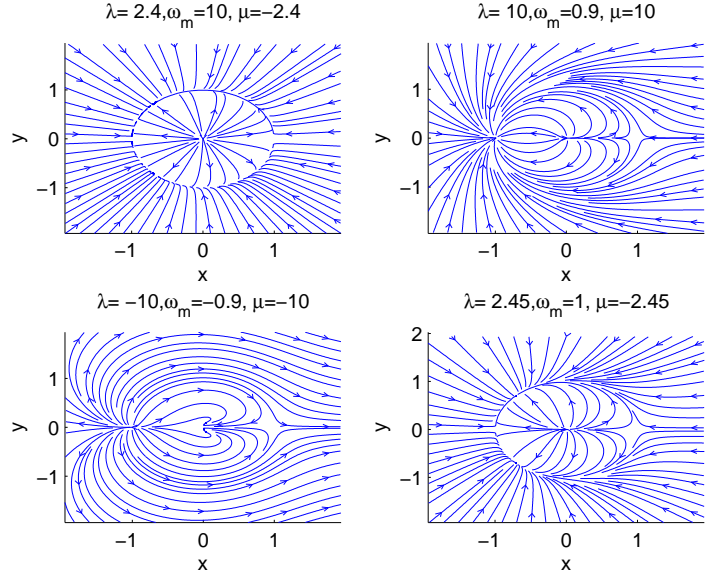}
\caption{Phase Plane Diagrams (I)}\label{fig111}
\end{center}
\end{figure}

\begin{figure}[H]
\begin{center}
\includegraphics
[scale=0.75]{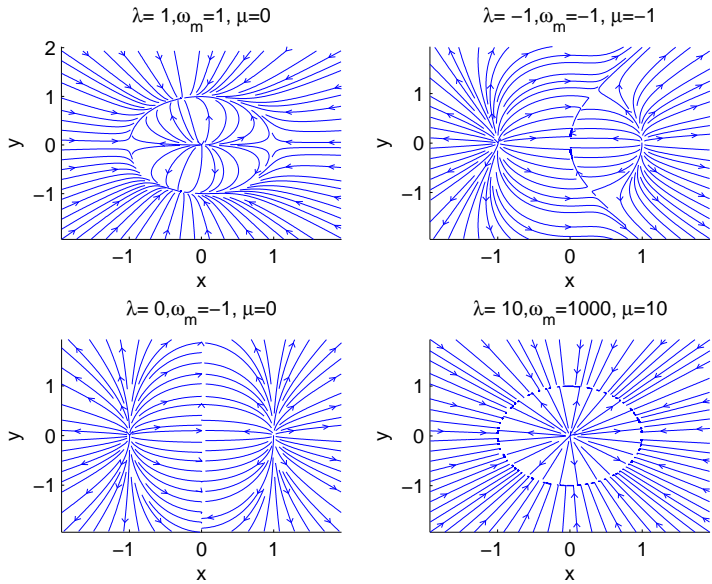}
\caption{Phase Plane Diagrams (II)}\label{fig112}
\end{center}
\end{figure}

\noindent We present the phase
plane diagrams of the autonomous system (\ref{ads})-(\ref{ads2}) for various
values of the parameter $\lambda, \omega_m$ and $\mu$'s in FIG. \ref{fig111} and FIG. \ref{fig112}. We note that, as all the critical points has the $\nu$ component $=1,$ the 2-D phase planes are actually the sections of solutions of (\ref{ads})-(\ref{ads2}) for particular values of $\lambda,\omega_m,\mu$  at $\nu=1.$

\smallskip

\smallskip
\bigskip

\bigskip

\noindent It is to be noted that the quintessence dark energy model with exponential potential can be regarded as a limiting case of   the DBI field with exponential potential and warp factor. A nice analytical treatment of autonomus systems for quintessence and other important models of dark energies can be found in~\cite{CST06} and some interesting applications of the center manifold theory for some of these dark energy models can be found in~\cite{BCL12}. Indeed, if $f(\phi)\dot{\phi}^2<<1,$ then from (\ref{lorentzlike}) it is clear that $\nu\approx 1.$ Therefore the field equations of the DBI field $\phi$ described in (\ref{dbi}) and the energy density (\ref{de.d}) and the pressure density (\ref{de.p}), all become identical to those of the quintessence field. Therefore it is no surprise that the critical points related to the quintessence model with the exponential potential, described in the the TABLE I in~\cite{CST06} should be $\nu\approx 1$ limit of the critical points related to the DBI field, described in the TABLE \ref{tab1} in this paper. There are five critical points, namely $(a),(b1),(b2),(c)$ and $(d)$ in the TABLE I in~\cite{CST06}. The critical point $(a)$ is a limiting case of the critical points $C_9$ and $C_{10}$ as $\omega_m\approx -1.$ The critical points $(b1),(b2)$ and $(c)$ are identical to the $C_1,C_2$ and $C_7$ respectively. The critical point $d$ is the $\nu\approx 1$ limit of $C_9.$ The critical points $C_3, C_4, C_5, C_6$ in our paper only pertain to the DBI field and do not arise in the analysis of the quintessence field. The critical point $(a)$ is ultrarelativistic in nature and $C_9, C_{10}$ only approaches it as $\omega_m\approx -1$, never coincides it. The results of a hyperbolic analysis has been listed in~\cite{CST06} for all these critical points in the TABLE I. The analysis of the non-hyperbolic cases for the critical point $(b1)$ and $(b2)$ has been done in the TABLES \ref{tab3} and \ref{tab5} respectively. In these tables, cases (b) and (d) pertain to the quintessence scenario and provide the result for the non-hyperbolic boundary cases.
 For the critical pont $(c),$ the center manifold analysis has been done in the TABLE \ref{tab16}. In the TABLE \ref{tab16}, only the case (b) relates to the quintessence situation. From the TABLE \ref{tab16}, it is evident that the stable nature of the non-hyperbolic critical point $C_7$ may correspond to the critical pont $(c)$ in~\cite{CST06}. The non-hyperbolic stability analysis for the critical point $C_9$ or it's $\nu\approx 1$ limit $(d),$ has been done in the TABLE \ref{tab22}. Only the case (b) is applicable for the quintessence case. It is clear that the stable nature of the non-hyperbolic critical point $C_9$ corresponds to the critical point (d). 

%%%%%%%%%%%%%%%%%%%%%%%%%%%%%%%%%%%%%%%%%%%%%%%%%%%%%%%%%%%%%%%%%%%%%%%%%%%%%%%%%%%%%%%%%%%%%%%%%%%%%%%%%%%%%%%%%%%%%%%
\section{Cosmological Interpretations and Conclusion}\label{secIV}
%%%%%%%%%%%%%%%%%%%%%%%%%%%%%%%%%%%%%%%%%%%%%%%%%%%%%%%%%%%%%%%%%%%%%%%%%%%%%%%%%%%%%%%%%%%%%%%%%%%%%%%%%%%%%%%%%%%%%%%
\noindent  The present work deals with the DBI-cosmology in the framework of dynamical system analysis. Here the DBI scalar field is chosen as the DE while a perfect fluid with barotropic equation of state is chosen as the dark matter. In the section \ref{secIII} we have seen that there are ten critical points $C_1-C_{10}.$ They are presented in the TABLE \ref{tab1}. The relevant cosmological parameters at these critical points are given in the TABLE \ref{tablast}. From these tables one may see that the pair of critical points $(C_1,C_2), (C_3,C_4), (C_5,C_6), (C_7,C_8)$ and $(C_9,C_{10})$ are cosmologically equivalent pairwise. We name these pairs as $A,B,C,D$ and $E$ respectively. If one considers the regions $E_{11}=\{\omega_m>1,\lambda<-\sqrt{6},\mu<\sqrt{6}\}$ and $E_{12}=\{\omega_m<1,\lambda>-\sqrt{6},\mu>\sqrt{6}\},$ then the stability analysis in the interiors of these two regions and in the interior of complement of $E_{11}\cup E_{12}$ is a simple application of the "Hartman-Gr\"{o}bman Theorem". The TABLE \ref{tab3} in this paper lists the different parts of the boundary between $(E_{11}\cup E_{12})$ and $(E_{11}\cup E_{12})^c$ (The superscript '$c$' denotes the complement) in case (a)-case (f). With the aid  of the Center Manifold Theory, the stability of the  system at these parts of the boundaries have been investigated and the results have been listed in the last column of the TABLE \ref{tab3}. From these results, we find that in all the other cases except for the part of the boundary described in case (f)$:=\{\omega_m\in\mathbf{R},\lambda=-\sqrt{6},\mu=\sqrt{6}\}$, $C_1$ presents a stable solution whereas on the part of the boundary (f) it is unstable. For the cosmologically equivalent critical point $C_2,$ it turns out that we have similar kind of results. Physically the pair of critical points $A$ represent a DBI scalar field dominated Universe where the scalar field is represented by a perfect fluid having the nature of a stiff fluid and correspond to a decelerating phase of the Universe. So it is not of much interest from the cosmological point of view. As an application of the Hartman-Gr\"{o}bman Theory, in~\cite{CSS10}, it has been shown that $C_3$ is stable on the region $E_{31}=\{\mu>\sqrt{3(1+\frac{1}{\omega_m})},\mu>-\lambda,\lambda<0\}$ and unstable on the region $E_{32}=\{\mu<\sqrt{3(1+\frac{1}{\omega_m})},\mu<-\lambda,\lambda<0\}$ and saddle on the interior of $(E_{31}\cup E_{32})^c.$ In this paper we have done the stability analysis on the different parts of the boundary between $(E_{31}\cup E_{32})$ and $(E_{31}\cup E_{32})^c$ with the aid of Center Manifold Theory. The different parts of the boundary and the stability results on them have been listed on the TABLE \ref{tab8}. These results are entirely new and can not be found in the literature such as ~\cite{CSS10}. $C_4$ has analogus set of results as $C_3.$ Physically the set of critical points $B$ are also scalar field dominated and it behaves as dark energy if $\mu^2<3$ (i.e. accelerating phase) while for $\mu^2>3,$ the DBI scalar field behaves as normal matter with the decelerating era of the Universe. For the critical point $C_5$ also, the stability analysis on the parts of the boundary between the open regions, determined by the negativity or the positivity of the eigenvalues has been done by the application of CMT and various stability results such as the stability of the system on the surface $\{\mu=-\lambda, \lambda\neq 0\}$ has been confirmed in this paper. As usual the critical point $C_6$ has entirely analogus set of results as $C_5$. Physically the pair of critical points $C$ represent scaling solution of the model where both the matter fields have contribution to the evolution. The Universe will be in the accelerating phase if $\omega_m<-\frac{1}{3}$ when the scalar field behaves as dark energy. The scaling solution will be dominated by the DE if $\omega_m<\frac{6}{\mu^2-6}$ otherwise it will be dominated by the dark matter. Similar is the situation for the critical points $E$ with DE dominance if $\omega_m>\frac{\lambda^2}{6}-1.$ Here for one of the critical points, say $C_{10},$ the parts of the boundary between the open regions in the $\omega_m-\lambda-\mu$ space as described above for the other critical points and the stability results on them has been described in the TABLE \ref{tab16}, \ref{tab19}, \ref{tab22} and \ref{tab25} with the aid of the CMT. These results are new. $C_9$ also has a similar set of results.
For $E_{71}=\{\frac{\lambda^2}{3}-1<\omega_m,\lambda>0,\mu>-\lambda\}$ and $E_{72}=\{\frac{\lambda^2}{3}-1<\omega_m,\lambda<0,\mu<-\lambda\}$, according to~\cite{CSS10}, $C_7$ is stable on $(E_{71}\cup E_{72})$. On $(E_{71}\cup E_{72})^c,$ it is unstable. However the boundary between the regions $(E_{71}\cup E_{72})$ and $(E_{71}\cup E_{72})^c,$ comprises of the surfaces described in the case (a), (b) and (c) of the TABLE \ref{tab14} in this paper, (a)$:=\{|\lambda|<\sqrt{6},\lambda\neq 0;\mu=-\lambda,\omega_m\in\mathbf{R}\}$, (b)$:=\{|\lambda|<\sqrt{6},\lambda\neq 0; \mu\in\mathbf{R},\omega_m=\frac{\lambda^2}{3}-1\}$ and (c)$:=\{|\lambda|<\sqrt{6},\lambda\neq 0;\mu=-\lambda,\omega_m=\frac{\lambda^2}{3}-1\}.$ By the application of CMT, on the surface (a) and (b), $C_7$ is stable whereas on the surface (c), $C_7$ is unstable. These results tabulated in the TABLE \ref{tab16} in our paper are entirely new. The stability results of the other critical point $C_8$ in the pair $D$ is analogus to $C_7.$ Physically the critical points set $D$ is dominated by the DE as the set $B$ with an accelerating phase for $\lambda^2<2$ and a decelerating phase for $\lambda^2>2.$ Therefore it can be inferred that from the cosmological viewpoint the critical points sets $C$ and $E$ are interesting and one may estimate the parameters $\mu$ and $\lambda$ from the observed results.

%%%%%%%%%%%%%%%%%%%%%%%%%%%%%%%%%%%%%%%%%%%%%%%%%%%%%%%%%%%%%

%%%%%%%%%%%%%%%%%%%%%%%%%%%%%%%%%%%%%%%%%%%%%%%%%%%%%%%%%%%%%

\end{document}